\documentclass[longauth]{aa} 

\usepackage{graphicx}
\usepackage{txfonts}
\usepackage{color}

\newcommand{\hls}{HLSJ091828.6+514223}
\usepackage{subfig}
\usepackage{threeparttable}
\usepackage[colorlinks=true, citecolor=blue]{hyperref}

\usepackage{natbib}
\bibpunct{(}{)}{;}{a}{}{,} 

\defcitealias{Bethermin+15}{B15}

\begin{document} 

\title{Faint millimeter NIKA2 dusty star-forming galaxies: finding the high-redshift population}

   \author{L.-J.~Bing \inst{\ref{LAM}} 
     \and A.~Beelen \inst{\ref{LAM}}
     \and  G.~Lagache \inst{\ref{LAM}}
     \and  R.~Adam \inst{\ref{LLR}}
     \and  P.~Ade \inst{\ref{Cardiff}}
     \and  H.~Ajeddig \inst{\ref{CEA}}
     \and  P.~Andr\'e \inst{\ref{CEA}}
     \and  E.~Artis \inst{\ref{LPSC}}
     \and  H.~Aussel \inst{\ref{CEA}}
     \and  A.~Beno\^it \inst{\ref{Neel}}
     \and  S.~Berta \inst{\ref{IRAMF}}
     \and  M.~Béthermin \inst{\ref{LAM}}
     \and  O.~Bourrion \inst{\ref{LPSC}}
     \and  M.~Calvo \inst{\ref{Neel}}
     \and  A.~Catalano \inst{\ref{LPSC}}
     \and  M.~De~Petris \inst{\ref{Roma}}
     \and  F.-X.~D\'esert \inst{\ref{IPAG}}
     \and  S.~Doyle \inst{\ref{Cardiff}}
     \and  E.~F.~C.~Driessen \inst{\ref{IRAMF}}
     \and  A.~Gomez \inst{\ref{CAB}} 
     \and  J.~Goupy \inst{\ref{Neel}}
     \and  F.~K\'eruzor\'e \inst{\ref{Argonne}}
     \and  C.~Kramer \inst{\ref{IRAMF},\ref{IRAME}}
     \and  B.~Ladjelate \inst{\ref{IRAME}} 
     \and  S.~Leclercq \inst{\ref{IRAMF}}
     \and  D.-Z.~Liu \inst{\ref{MPE}}
     \and  J.-F.~Lestrade \inst{\ref{LERMA}}
     \and  J.~F.~Mac\'ias-P\'erez \inst{\ref{LPSC}}
     \and  A.~Maury \inst{\ref{CEA}}
     \and  P.~Mauskopf \inst{\ref{Cardiff},\ref{Arizona}}
     \and  F.~Mayet \inst{\ref{LPSC}}
     \and  A.~Monfardini \inst{\ref{Neel}}
     \and  M.~Mu\~noz-Echeverr\'ia \inst{\ref{LPSC}}
     \and  R.~Neri \inst{\ref{IRAMF}}
     \and  L.~Perotto \inst{\ref{LPSC}}
     \and  G.~Pisano \inst{\ref{Roma}}
     \and  N.~Ponthieu \inst{\ref{IPAG}}
     \and  V.~Rev\'eret \inst{\ref{CEA}}
     \and  A.~J.~Rigby \inst{\ref{Cardiff}}
     \and  A.~Ritacco \inst{\ref{ENS}, \ref{INAF}}
     \and  C.~Romero \inst{\ref{Pennsylvanie}}
     \and  H.~Roussel \inst{\ref{IAP}}
     \and  F.~Ruppin \inst{\ref{IP2I}}
     \and  K.~Schuster \inst{\ref{IRAMF}}
     \and  S.~Shu \inst{\ref{Caltech}} 
     \and  A.~Sievers \inst{\ref{IRAME}}
     \and  C.~Tucker \inst{\ref{Cardiff}}
     \and  M.-Y.~Xiao \inst{\ref{CEA}, \ref{NJU}}
     \and  R.~Zylka \inst{\ref{IRAMF}}
     }
   \institute{Aix Marseille Univ, CNRS, CNES, LAM (Laboratoire d'Astrophysique de Marseille), Marseille, France
     \label{LAM}
     \and 
     LLR (Laboratoire Leprince-Ringuet), CNRS, École Polytechnique, Institut Polytechnique de Paris, Palaiseau, France
     \label{LLR}
     \and
     School of Physics and Astronomy, Cardiff University, Queen’s Buildings, The Parade, Cardiff, CF24 3AA, UK 
     \label{Cardiff}
     \and
     AIM, CEA, CNRS, Universit\'e Paris-Saclay, Universit\'e Paris Diderot, Sorbonne Paris Cit\'e, 91191 Gif-sur-Yvette, France
     \label{CEA}
     \and
     Univ. Grenoble Alpes, CNRS, Grenoble INP, LPSC-IN2P3, 53, avenue des Martyrs, 38000 Grenoble, France
     \label{LPSC}
     \and
     Institut N\'eel, CNRS, Universit\'e Grenoble Alpes, France
     \label{Neel}
     \and
     Institut de RadioAstronomie Millim\'etrique (IRAM), Grenoble, France
     \label{IRAMF}
     \and 
     Dipartimento di Fisica, Sapienza Universit\`a di Roma, Piazzale Aldo Moro 5, I-00185 Roma, Italy
     \label{Roma}
     \and
     Univ. Grenoble Alpes, CNRS, IPAG, 38000 Grenoble, France 
     \label{IPAG}
     \and
     Centro de Astrobiolog\'ia (CSIC-INTA), Torrej\'on de Ardoz, 28850 Madrid, Spain
     \label{CAB}
     \and
     High Energy Physics Division, Argonne National Laboratory, 9700 South Cass Avenue, Lemont, IL 60439, USA
     \label{Argonne}
     \and  
     Instituto de Radioastronom\'ia Milim\'etrica (IRAM), Granada, Spain
     \label{IRAME}
     \and
     Max-Planck-Institut für Extraterrestrische Physik (MPE), Giessenbachstr. 1, D-85748 Garching, Germany
     \label{MPE}
     \and
     LERMA, Observatoire de Paris, PSL Research University, CNRS, Sorbonne Universit\'e, UPMC, 75014 Paris, France  
     \label{LERMA}
     \and
     Department of Physics and Astronomy, University of Pennsylvania, 209 South 33rd Street, Philadelphia, PA, 19104, USA
     \label{Pennsylvanie}
     Laboratoire de Physique de l’\'Ecole Normale Sup\'erieure, ENS, PSL Research University, CNRS, Sorbonne Universit\'e, Universit\'e de Paris, 75005 Paris, France 
     \label{ENS}
     \and
     INAF-Osservatorio Astronomico di Cagliari, Via della Scienza 5, 09047 Selargius, IT
     \label{INAF}
     \and 
     Institut d'Astrophysique de Paris, CNRS (UMR7095), 98 bis boulevard Arago, 75014 Paris, France
     \label{IAP}
     \and
     University of Lyon, UCB Lyon 1, CNRS/IN2P3, IP2I, 69622 Villeurbanne, France
     \label{IP2I}
     \and
     School of Earth and Space Exploration and Department of Physics, Arizona State University, Tempe, AZ 85287, USA
     \label{Arizona}
     \and
     Caltech, Pasadena, CA 91125, USA
     \label{Caltech}
     \and
     School of Astronomy and Space Science, Nanjing University,
     Nanjing 210093, China
     \label{NJU}
   }

   \date{Received ; accepted }

 
  \abstract
   {} 
   {High redshift dusty star-forming galaxies are proposed to be the progenitors of massive quiescent galaxies arising in the cosmic noon, providing a crucial insight on the formation, assembly, and early quenching of massive galaxies in the early Universe. However, their high redshift combined with high dust obscuration adds significant difficulties to their redshift measurement, which is mandatory for detailed studies on their physical properties. Blind millimeter spectral scans are in principle the most unbiased way to obtain accurate spectroscopic redshifts for these sources, while they also suffer from the difficulty of identifying faint molecular/atomic lines within limited telescope time for faint DSFGs.} 
   {We develop a new framework to constrain the source redshift. The method jointly accounts for the detection/non-detection of spectral lines and the prior information from the photometric redshift and total infrared luminosity from spectral energy distribution analysis. The method uses the estimated total infrared luminosity to predict the line fluxes at given redshifts and generates model spectra. The redshift-dependent spectral models are then compared with the observed spectra to find the redshift.} 
   {We apply the aforementioned joint redshift analysis method to four high-z dusty star-forming galaxy candidates selected from the NIKA2 observations of the \hls\, (HLS) field, and further observed by NOEMA with blind spectral scans. These sources only have SPIRE/Herschel photometry as ancillary data. They were selected because of very faint or no SPIRE counterparts, as to bias the sample towards the highest redshift candidates. The method finds the spectroscopic redshift of 4 in the 5 NOEMA-counterpart detected sources, with z>3.  
   Based on these measurements, we derive the CO/[CI] lines and millimeter continuum fluxes from the NOEMA data and study their ISM and star-formation properties. We find cold dust temperatures in some of the HLS sources compared to the general population of sub-millimeter galaxies, which might be related to the bias introduced by the SPIRE-dropout selection. Our sources, but one, have short gas depletion time of a few hundred Myrs, which is typical among high-z sub-millimeter galaxies. The only exception shows a longer gas depletion time, up to a few Gyrs, comparable to that of main-sequence galaxies at the same redshift. Furthermore, we identify a possible over-density of dusty star-forming galaxies at z=5.2, traced by two sources in our sample, as well as the lensed galaxy \hls.} 
  {We demonstrate that our method applied to millimeter-selected DSFGs could determine the redshift accurately. Such accuracy with only multiple low S/N emission lines shows promising potential for the blind redshift search on large samples of high-z DSFGs, even in the absence of optical-IR photometric redshifts.}


   \maketitle

\section{Introduction}
It is now clearly established that dusty star-forming galaxies (DSFGs) are critical players in the assembly of galaxy stellar mass and the evolution of massive galaxies at $z<3$ \citep[e.g.][]{Madau+14}. At higher redshift, observing the dusty star-formation and its spatial and redshift distribution requires undoubtedly (sub-)mm experiments and is still very challenging. For example, the limited existing estimates on dust-obscured star formation rate densities (SFRD) at $z>4$ are still not consistently measured, as shown in the discrepancy between recent studies \citep[e.g.,][]{Gruppioni+20, Dudzeviciute+20, Fudamoto+21,Zavala+21, fujimoto+23}. This is largely due to difficulties in uncovering a large unbiased sample of high-redshift DSFGs in relatively large cosmic volumes. Bright and faint DSFGs at high redshift have been uncovered by SPT \citep{Reuter+20} and ALMA surveys \citep{Franco+18, Zavala+21, Aravena+20}. However, statistical studies with these sources suffer from the fact that either strongly-lensed DSFG samples are not well statistically defined or covered areas are limited. \\

It is well known that in the (sub-)millimeter, larger area and relatively deep surveys can efficiently find high-redshift DSFGs \citep{Bethermin+15a}, thanks to the negative k-correction \citep[e.g.,][]{Casey+14}, combined with the shape of the luminosity functions. Such large-area deep surveys are conducted with single-dish telescopes, as with the SCUBA2 instrument on the JCMT \citep{Holland+13} or the NIKA2 instrument on the IRAM 30m \citep{Perotto+20}. The angular resolutions of such single-dish surveys are 13", 11.1” and 17.6”, for SCUBA2 at 850\,$\mu$m, and NIKA2 at 1.2 and 2\,mm respectively. This makes it difficult to unambiguously identify the multi-wavelength counterparts of the DSFGs and to search for the high-redshift population. As already shown by the follow-ups of SCUBA2 sources with ALMA \citep[e.g.,][]{Simpson+20}, the combination of single-dish and interferometer surveys is by far the most efficient way of constraining the dusty star formation at $2 < z < 6$. Indeed, the high resolution and sensitivity of (sub-)millimeter interferometers can provide accurate position measurements on DSFGs and thus the identification of their multi-wavelength counterparts. However, getting photometric redshift from optical-IR is complicated by the lack of sufficiently deep homogeneous multi-wavelength data to analyze large samples. Moreover, DSFGs are subject to significant optical extinction (some of them are even optically dark, see \citealt{Franco+18, Williams+19, Manning+21}) which impacts the quality and reliability of photometric redshift estimates and prevents optical/near-infrared spectroscopic follow-up. Photometric redshifts from far-IR/mm to radio broad-band photometry have been used in studies on the cosmic evolution of high-z DSFGs since the discovery of DSFGs \citep{Yun+02,Hughes+02,Negrello+10}. However, these measurements are even more uncertain than the optical-IR photometric redshift, as the spectral energy distributions (SEDs) in the far-IR/mm do not show any spectral features (but a broad peak), and there are often only few data points on the SEDs to constrain the model. In addition, there is a strong degeneracy between dust temperature and redshift in distant dusty galaxy, which limits the usefulness of simple photometric redshifts \citep[e.g.,][]{Blain99}.  Finally, in the modelling of the FIR emission, 
optically thin or thick solutions are heavily degenerate. Indeed, the same SED could arise from either cold and optically thin or from a warmer and optically thicker FIR dust emission with no robust way to discriminate between the two by using continuum observations \citep{Cortzen+20}. This often leads to an overestimate of the FIR photometric redshifts because of an apparent colder dust temperature derived from optically-thin emission in high-redshift, starbursting DSFGs \citep{Jin+19}.

For such galaxies, spectral scans in the millimeter can be the only way of getting the spectroscopic redshift, as shown in e.g. \cite{Walter+12, Fudamoto+17, Strandet+17, Zavala+18}. The success rate of measuring the redshift using millimeter spectral scans can be very high, being $>$70\% \citep{Weiss+13, Strandet+16} and even up to $>$90\% \citep{Neri+20}. Such a success rate is obtained on large samples in a reasonable amount of telescope time but for bright DSFGs. For example, with a total time of 22.8\,hours on 13 DSFGs with average 850$\rm \mu$m fluxes of 32\,mJy, \cite{Neri+20} measured the redshift of 12/13 sources with NOEMA. \cite{Weiss+13} obtained a $\sim$90\% detection rate for sources with $S_{1.4\,mm}>20$\,mJy. Obviously, for much fainter objects, obtaining redshifts may become much more difficult \citep[e.g.][]{Jin+19}.

We are currently conducting a deep survey with NIKA2, the NIKA2 Cosmological Legacy Survey (N2CLS), a guaranteed time observation (GTO) large program searching for a large sample of high-z DSFGs \citep{Bing+21, Bing+23}. The observations cover two fields, GOODS-N and COSMOS, and most of the detected DSFGs are sub-mJy sources at 1.2\,mm. One of the goals of N2CLS is to put new solid constraints on the obscured SFRD at $z>4$. To reach that goal, we need first to obtain the redshift of N2CLS sources. While deep optical-IR data are available in the two fields and have been extensively used to obtain photometric redshifts, a large fraction of the sources currently lack a secure redshift. Given the wealth of ancillary data already available on these two fields, blind millimeter spectral scans is the only solution to measure their spectroscopic redshift. As a pilot program to try to identify the high-redshift population, we selected 4 high-redshift candidates detected at 1.2 and 2\,mm by NIKA2. They have been selected from their far-IR/mm SEDs photometric redshift in the \hls\, field observed with NIKA2 during the Science Verification. This paper presents the redshift identification and source properties based on the spectral scans obtained with NOEMA on these sources. It is organised as follows. In Sect\,\ref{sc:sample_selection}, we present the sample and NIKA2 observations. Section\,\ref{sc:data_reduction} describes the NOEMA observations and data reduction, as well as the extraction of continuum fluxes and spectral scans. In Sect.\,\ref{sc:redshift_analysis}, we extensively discuss the redshifts. In particular, we develop a new method that combines both far-IR to millimeter photometric data and spectral scans to measure the redshift. Source properties, as their dust mass and temperature, kinematics and excitation of molecular gas, are given in Sect.\,\ref{sc:properties}. Section\,\ref{sc:overdensity} presents the potential discovery of a DSFG over-density at z=5.2 in the HLS field. Conclusions on the main results and the possible implications of our findings in future high-z DSFGs studies are given in Sect.\,\ref{sc:conclusions}. Finally, three appendices give more details on the method of redshift measurements and its validation. Throughout the paper, we adopt the standard flat $\Lambda$CDM model as our fiducial cosmology, with cosmological parameters H$_0$=67.7 km/s/Mpc, $\Omega_m$=0.31 and $\Omega_\Lambda$=0.69, as given by \cite{Planck+18}.

\section{Sample selection and NIKA2 observations}\label{sc:sample_selection}

\begin{table*}
\caption{Coordinates, millimeter fluxes and SPIRE far-IR fluxes of our NIKA2 (HLS) sample. }

\label{sourceinfo}
\centering
\begin{tabular}{cccccccc}
\hline\hline
Source & RA & Dec & F$_{SPIRE250}$ & F$_{SPIRE350}$ & F$_{SPIRE500}$ & F$_{NIKA2-1.2}$ & F$_{NIKA2-2.0}$ \\
 & & & mJy & mJy & mJy & mJy & mJy \\
\hline
 HLS-2 & 09:18:17.2 & 51:41:25.1 & (6.1) & 11.3$\pm$6.1 & 17.4$\pm$6.1 & 2.9$\pm$0.3 & 0.42$\pm$0.07 \\
 HLS-3 & 09:18:23.3 & 51:42:51.9 & (6.1) & (6.1) & (6.3) & 2.4$\pm$0.3 & 0.60$\pm$0.06 \\
 HLS-4 & 09:18:24.3 & 51:40:49.7 & (6.1) & 8.9$\pm$6.1 & 8.9$\pm$6.1 & 1.9$\pm$0.3 & 0.28$\pm$0.07 \\
 HLS-22 & 09:18:34.9 & 51:41:44.9 & (6.1) & (6.1) & (6.1) & 1.7$\pm$0.3 & 0.36$\pm$0.06 \\
 \hline
\end{tabular}

\end{table*}

\subsection{NIKA2 field around \hls}
As part of the NIKA2 Science Verification that took place in February 2017, we observed an area of 185~arcmin$^2$, centered on \hls\,, a lensed dusty galaxy at z=5.24 \citep{Combes+12}, for a on source time of about 3.5\,hours at the center.
This allowed us to reach $1\sigma$ sensitivities of about 0.3\,mJy at 1.2\,mm and 0.1\,mJy at 2\,mm on \hls. This galaxy is close to the z=0.22 cluster Abell\,773, but is likely lensed by a galaxy at z$\sim$0.63. For our NIKA2 sources, the magnification by the galaxy cluster is $<$10$\%$. Therefore, we do not expect the NIKA2 sources to be highly magnified (E. Jullo, private communication).

The NIKA2 field overlaps almost entirely with Herschel SPIRE observations at 250, 350, 500\,$\mu$m. On the contrary, the PACS, IRAC and HST images cover only a very small part of the field on the west side (where NIKA2 observations have lower signal-to-noise ratios). Thus only SPIRE data were used to select high-z candidates.
The SPIRE fluxes were measured using FASTPHOT\footnote{https://www.ias.u-psud.fr/irgalaxies/downloads.php} \citep{Bethermin+10} through simultaneous PSF fitting, using NIKA2 source positions as priors on the SPIRE maps.

We built a 1.2 and 2\,mm catalog using the NIKA2 data reduced using the collaboration pipeline (Ponthieu et al., in prep). A total of 27 sources are detected with S/N$>$5 in at least one band (1.2 or 2\,mm). From this catalog, we selected four sources detected at both 1.2 and 2\,mm with high signal to noise ratio (between 5.7 and 9.7) and for which there is a faint (at the level of confusion noise) or no SPIRE counterparts, as to bias the sample towards the highest redshift candidates. Indeed, rough sub-millimeter photometric redshifts, obtained by fitting empirical IR SED templates from \cite{Bethermin+15} to our SPIRE+NIKA2 data, were $z_{phot}\sim5-7$. These sources are named HLS-2, HLS-3, HLS-4, and HLS-22. Their fluxes are between 1.7 and 2.9\,mJy at 1.2\,mm and 0.28 and 0.60\,mJy at 2\,mm. The flux measurements and uncertainties are presented in Table~\ref{sourceinfo}, where the 1-sigma flux uncertainty of SPIRE undetected HLS sources are in parenthesis. The quoted uncertainties account only for uncertainties coming from flux measurements.

     \begin{table*}
  \caption{Information on NOEMA follow-up observations.}              
  \label{noema_info}      
  \centering                                      
  \begin{tabular}{c c c c c c}          
  \hline\hline                        
  Source Name & Setup & $\nu$ [GHz] & $\sigma_{cont}$ [$\mu$Jy/beam] & $\sigma_{channel,2MHz}$ [mJy/beam] & On-source time [h] \\    
  \hline                                   
    HLS-2 & S20CL001 & 137.5 & 16 & 0.1 & 12.0 \\
       & W17EL002 & 84.25 & 24 & 0.4 & 0.6 \\   
       & W17EL001 & 76.50 & 13 & 0.2 & 4.4 \\   
    HLS-3 & S20CL002 & 144.2 & 21 & 0.3 & 2.4 \\
       & W17EL002 & 84.25 & 23 & 0.4 & 0.6 \\ 
       & W17EL001 & 76.50 & 15 & 0.2 & 4.5 \\      
    HLS-4 & W17EL002 & 84.25 & 22 & 0.4 & 0.6 \\ 
       & W17EL001 & 76.50 & 13 & 0.4 & 4.5 \\
    HLS-22 & W18FA001 & 114.0 & 8 & 0.2 & 5.2 \\
       & W17FA002 & 84.25 & 12 & 0.2 & 5.4 \\
       & W17FA001 & 76.50 & 15 & 0.3 & 4.7 \\
             
  \hline                                             
  \end{tabular}
  \end{table*}
  
\section{NOEMA Observations and source identification}\label{sc:data_reduction}

\subsection{NOEMA observations and data calibration}\label{sc:calib}
   
   Follow-up observations were made using NOEMA from 2018 to 2020, with 4 different programs. The 4 sources in the HLS field were all observed by NOEMA with the PolyFiX correlator. They were initially targeted by project W17EL (HLS-2/3/4) 
   and W17FA (HLS-22) 
   using the same setups that continuously cover the spectra from 71\,GHz to 102\,GHz with the D configuration in band1. HLS-22 were further observed in project W18FA 
   with the A configuration in band1 and HLS-2/3 were further observed in project S20CL 
   with the D/C configuration in band2. The total on-source time of all of the proposals is 44.9\,hours. The details of the observations on each source are summarized in Table~\ref{noema_info}. 

   NOEMA observations are first calibrated using CLIC and imaged by MAPPING under GILDAS\footnote{https://www.iram.fr/IRAMFR/GILDAS}. Radio sources 3C454.3, 0716+714, 1156+295, 1055+018, 0851+202 and 0355+508 are used for bandpass calibrations during these observations, and the source fluxes are calibrated using LHKA+101 and MWC349. With the calibrated data, we further generate the uv table with the original resolution of 2\,MHz. We also produce the continuum uv table of each source by directly compressing all corresponding lower sideband (LSB) and upper sideband (USB) data with the uv\_compress function in MAPPING. 

   \subsection{NOEMA continuum flux measurement and source identification}\label{sc:continfo}
   
\begin{figure*}[tb]
   \centering
      \includegraphics[clip=true,width=.99\textwidth]{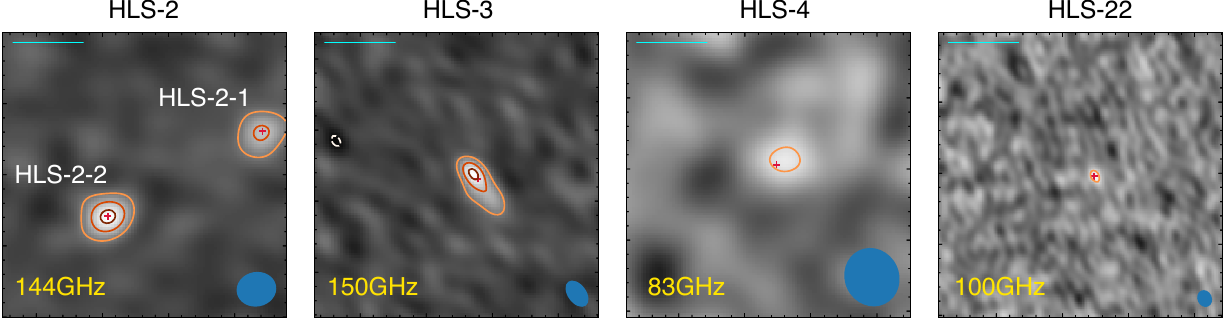}
      \caption{Cleaned images of NOEMA observation on our four NIKA2 sources. The effective beam size and shape of each map is shown in the bottom right of each panel. The contour levels from orange to dark red correspond to -4, 4, 8 and 12$\times$ RMS of each map, respectively. The red crosses mark the position of detected NOEMA sources from the uv\_fit. The two resolved sources associated with HLS-2 are also labeled separately (HLS-2-1 and HLS-2-2). The scale bars in the maps (upper left) correspond to 5 arcseconds in the sky. The frequency of the continuum data are given in the lower left corner of each panel.}
      \label{noemaclean}
   \end{figure*}

   We identify the counterparts of our sample in the NOEMA continuum data.
   We first generate the continuum dirty map and then clean the continuum image of each source with the Clark algorithm within MAPPING.
   The cleaned image of each source with the highest SNR and(or) the best spatial resolution is shown in Fig.~\ref{noemaclean}. We blindly search for candidate sources by identifying all of the peaks above 4$\times$RMS within the NOEMA primary beam. Their accurate positions are then derived with uv\_fit function in MAPPING (with the peak positions as the initial prior and point source as the model), and the continuum fluxes at the other frequencies are estimated with source models fixed to these reference images. 
   
         The continuum fluxes are measured using uv\_fit and the same models as given in Table\,\ref{cont_noema_pos_shape}. The 4 sidebands in the 2 setups of W17EL and W17FA are combined together to generate continuum uv tables centered on 3.6\,mm, given the low SNR of the continuum emission at such a long wavelength. When data are available, the continuum fluxes at higher frequencies are measured both sideband by sideband and on the combined LSB+USB uv-table.  
   The continuum fluxes are listed in Table\,\ref{noema_cont_flux}.
   
     We detect 5 reliable continuum sources within the primary beam of NOEMA observations as counterparts of 4 NIKA2 HLS sources. We further checked the residual RMS on the map with source models more complex than point sources. This does not improve the level of residuals of three NOEMA sources. For the rest two sources, HLS-2-1 and HLS-3, the favored simplest models are a circular Gaussian model and an elliptical Gaussian model, respectively. 
   The position and preferred models of each source is listed in Table~\ref{cont_noema_pos_shape}, and we note that the position of these sources does not change significantly depending on the model.

   We show on Fig.\,\ref{noemaclean} the cleaned images of NOEMA observations. The NIKA2 source HLS-2 is resolved into two continuum sources in our high-resolution NOEMA observation with SNR$\sim$10. The rest of NIKA2 sources are all associated with one single NOEMA source. For these sources (HLS-3, HLS-4 and HLS-22), we compare their positions in the NOEMA and NIKA2 observations. The maximum offset is found in HLS-3 with a value of 1.9\,arcsec. The average offset is 0.9\,arcsec among these three sources, which suggests a high positional accuracy of NIKA/NIKA2 for locating sources with relatively high SNR.

\begin{table}
\caption{NOEMA continuum source positions and best-fit sizes.}              
\label{cont_noema_pos_shape}      
\centering                                      
\begin{tabular}{c c c c}          
\hline\hline                        
Name & RA & Dec & FWHM(") \\    
\hline                                   
   HLS-2-1  & 09:18:16.3 & 51:41:28.1 & 1.2  \\
   HLS-2-2  & 09:18:17.5 & 51:41:22.1 & point \\  
   HLS-3  & 09:18:23.1 & 51:42:51.6 & 2.8$\times$0.7  \\   
   HLS-4  & 09:18:24.26 & 51:40:50.3 & point  \\   
   HLS-22  & 09:18:34.76 & 51:41:44.8 & point  \\   
\hline                                             
\end{tabular}
\end{table}

\begin{table}
\caption{Continuum fluxes from NOEMA observations.}              
\label{noema_cont_flux}      
\centering                                      
\begin{tabular}{c c c c c}          
\hline\hline                        
Name & Sideband &$\nu_{cont}$ [GHz] &  S$_{cont}$ [$\mu$Jy] & SNR \\    
\hline                                   
   HLS-2-1 & LSB+USB & 143.7 & 239$\pm$30\tablefootmark{*} & 8.1 \\  
    -- & LSB+USB &82.5 & 42$\pm$15 & 2.7 \\   
   HLS-2-2 & LSB+USB & 143.7 & 231$\pm$13\tablefootmark{*} & 17.9 \\
    -- & LSB+USB & 82.5 & 44$\pm$13 & 3.5 \\ 
   HLS-3 & LSB+USB & 150.0 & 418$\pm$44\tablefootmark{*} & 9.6 \\    
    -- & LSB+USB & 82.5 & 32$\pm$14 & 2.3 \\
   HLS-4 & LSB+USB & 82.5 & 48$\pm$10\tablefootmark{*} & 4.8 \\   
   HLS-22 & USB & 113.7 & 58$\pm$16 & 3.6 \\    
    -- & LSB & 100.3 & 45$\pm$10\tablefootmark{*} & 4.5 \\ 
    -- & LSB+USB & 81.9 & <42 & N/A\\           
\hline                                            
\end{tabular}
\tablefoot{\tablefoottext{*}{Data sets and fluxes derived with free parameters on source position and shape in uv\_fit. The fluxes at the other frequencies on a specific source are fitted with positions and shapes fixed to the same as marked data set, which are given in Table~\ref{cont_noema_pos_shape}.}}
\end{table}

   For HLS-2 and HLS-3, part of our NOEMA observations measure their continuum fluxes at a frequency close to the representative frequencies of NIKA2 2\,mm bands. The NOEMA and NIKA2 fluxes are consistent for HLS-3. The total NOEMA fluxes of the 2 components of HLS-2 is 50\% higher than that measured by NIKA2, while still being consistent with each other within 3\,$\sigma$ uncertainties.  This first comparison is encouraging. A detailed study of NIKA2 and NOEMA fluxes is beyond the scope of this paper and will be conducted with more statistics (e.g. with the NOEMA follow-up of N2CLS sources).    
   
   \subsection{Extraction of NOEMA millimeter spectra}\label{sc:specext}
   
   We extract the millimeter spectra of NOEMA continuum sources from the full uv table. The uv tables are first compressed by the uv\_compress function in MAPPING, which makes averages within several channels to enhance the efficiency of the line searching with higher SNR per channel and smaller load of data. For observations in band1 we set the number of channels to average to 15 while the observations in band2 and band3 are averaged every 25 channels, which corresponds to channel widths of 107km/s, 100km/s and 59km/s at 84\,GHz, 150\,GHz and 255\,GHz, respectively. Given the typical line width (a few hundred to one thousand km/s) for sub-millimeter galaxies \citep{Spilker+14}, the compression of uv tables could still ensure Nyquist sampling by 2-3 channels on the emission line profiles and preserve the accuracy of line center and redshift measurement. 
   
   To extract the spectra, we perform uv\_fit on the compressed spectral uv table with the position and source model fixed to the same as those given in Table\,\ref{cont_noema_pos_shape}. For the observations of the W17EL002 setup, we flagged the visibilities associated with one antenna significantly deviating from the others. Given the relatively low angular resolution of most of our data on HLS sources ($\sim$5" in band1 and $\sim$2" in band2), these galaxies is unlikely to be significantly resolved, thus the uv\_fit at fixed position on the uv tables should be able to uncover the majority of their line emission. 
   
\begin{figure*}[tb]
   \centering
      \includegraphics[clip=true,width=.99\textwidth]{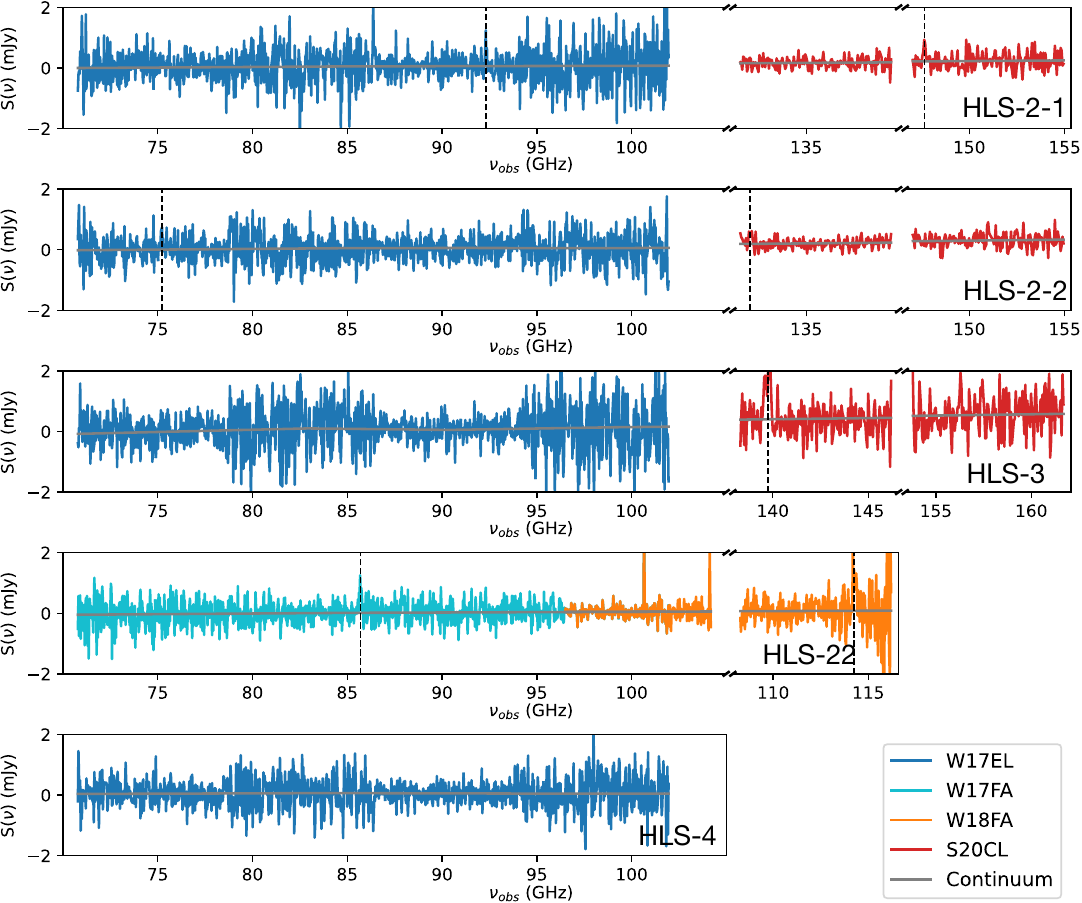}
      \caption{Millimeter spectra of all HLS sources extracted from the uv tables obtained from the NOEMA observations. The continuum models to be subtracted are presented as grey solid lines. The lines identified to determine the spectroscopic redshift of the sources (see Sect.~\ref{sc:redshift_analysis} and Fig.~\ref{joint_redshift_HLS} for details) are marked  by dashed black vertical lines.}
      \label{plot_blindext}
   \end{figure*}

   We further remove the continuum in the extracted spectra, assuming a fixed spectral index of 4. This is equivalent to a modified black-body spectrum with a fixed emissivity ($\beta$) at 2, and is generally consistent with the dust emissivity we derived in Sect.~\ref{sc:properties_dust_mbb}. We use these continuum subtracted NOEMA spectra for the redshift search and the emission line flux measurement (see Sect.~\ref{sc:joint_pdf_z} and Sect.~\ref{sc:properties_gas_hls}). The extracted spectra and continuum model to be removed are shown in Fig.~\ref{plot_blindext}.
   
   \section{Source redshift from photometric-spectroscopic joint analysis}\label{sc:redshift_analysis}
   
   An accurate redshift is a prerequisite to the accurate estimate of the physical properties of high-z galaxies. However, the optical-IR SED of high-z DSFGs are often much poorer constrained than other high-z galaxies due to their faintness at these wavelengths, which poses challenges to the accurate measurement of their photometric redshifts. In far-IR, the degeneracy between colors, dust temperature and redshift could also lead to a highly model dependent estimate of the photometric redshift. In this section, we describe the different methods and summarize the results of redshift estimate on our sample, with both photometric and spectroscopic data described in Sect.\,\ref{sc:data_reduction}. Specifically, we introduce a new joint analysis framework to determine the redshifts of NIKA2 sources combining the probability distribution function of photometric redshifts together with the corresponding IR luminosities and blind spectral scans, which helps us identify the low SNR spectral lines in the NOEMA spectra.

\begin{table}
\caption{Photometric redshifts of HLS sources.}              
\label{photoz_HLS}      
\centering                                      
\begin{tabular}{c c c c} 
\hline\hline                        
Name &  z$_{B15, ms}$ & z$_{B15, sb}$ & z$_{MMPZ}$\\    
\hline                          
   HLS-2  & 4.6$^{+0.8}_{-0.7}$ & 4.3$^{+0.7}_{-0.7}$ & 3.5$^{+1.5}_{-0.8}$ \\
   HLS-3  & 5.9$^{+1.2}_{-0.9}$ & 5.7$^{+1.1}_{-0.9}$ & 4.6$^{+2.5}_{-1.4}$ \\
   HLS-4  & 4.3$^{+1.0}_{-1.0}$ & 4.1$^{+1.0}_{-0.8}$ & 3.3$^{+0.6}_{0.6}$ \\
   HLS-22 & 4.7$^{+1.3}_{-1.0}$ & 4.5$^{+1.2}_{-0.9}$ & 3.1$^{+4.6}_{-1.4}$ \\
   
\hline                                      
\end{tabular}
\end{table}
   
   \subsection{Photometric redshifts}\label{sc:photo-z_hls}
   
  The lack of deep optical and infrared data for the HLS field makes it impossible to conduct a full SED modeling of NIKA2 detected sources. However, with the NIKA2 and SPIRE photometry, we fit the far-IR SED of HLS sources with dust emission templates to estimate their redshifts and IR luminosities. Given the poor angular resolution of the FIR data, we are not able to obtain the fluxes of each single component resolved by NOEMA observations on HLS-2. Thus we only fit with the integrated fluxes under the assumption that the 2 components blended within the beam of SPIRE and NIKA2 are located at the same redshift. 
   
   We used 2 sets of FIR dust templates: the synthetic infrared SED templates from \citet{Bethermin+15} (herafter \citetalias{Bethermin+15}) and the MMPZ framework \citep{Casey+20} using parametrized dust templates from \citet{Casey+12}. 
   
   \citetalias{Bethermin+15} templates could be described as a series of empirical dust SEDs of galaxies at different redshifts. The dust SEDs are produced based on the deep observational data from infrared to millimeter. It considers 2 populations of star-forming galaxies, starburst and main-sequence galaxies, and produces the 2 sets of empirical SED templates correspondingly. 
   We fit our photometric data points with the templates of main-sequence galaxies, which consist of 13 SEDs at each redshift. These templates include the average SED and the SEDs within $\pm$3$\sigma$ uncertainties with steps of 0.5$\sigma$. The estimated redshift, as well as the 1$\sigma$ uncertainties based on the fitting using \citetalias{Bethermin+15} main sequence and starburst SED templates are listed in Table\,\ref{photoz_HLS}. In the following redshift searching involving \citetalias{Bethermin+15} templates (see Sect.~\ref{sc:joint_pdf_z} and Sect.~\ref{sc:z_results}), we will only use and present the output of SED fitting based on main sequence templates. This is mainly because the output results of the SED fitting based on starburst and main sequence templates, as shown in Table\,\ref{photoz_HLS}, are highly consistent considering the uncertainty.
  
   \citet{Casey+12} describes the intrinsic FIR dust emission of galaxies using a generalized modified black-body model in far-IR plus a power-law model at mid-IR. For the SED fitting with the \citet{Casey+12} template, we work within the framework of the MMPZ algorithm \citet{Casey+20}. It considers the intrinsic variation of dust SED at different IR luminosities, as well as the impact of the rising CMB temperature at high redshift. The default set of IR SEDs fixes the mid-infrared spectral slope to 3 and dust emissivity $\beta$ to 1.8. The template SED also considers the transition from optically thin to optically thick when going to lower wavelengths, where the wavelength of unity opacity ($\tau(\lambda)$=1) is fixed to 200\,$\mu$m. The redshift, the total infrared luminosity and the corresponding wavelength at the peak of IR SED are the main parameters to be considered for the fit. The empirical correlation between the latter two parameters is also taken into account during the fit.
   
   \begin{figure*}[!ht]
   \centering
      \includegraphics[width=1.0\textwidth]{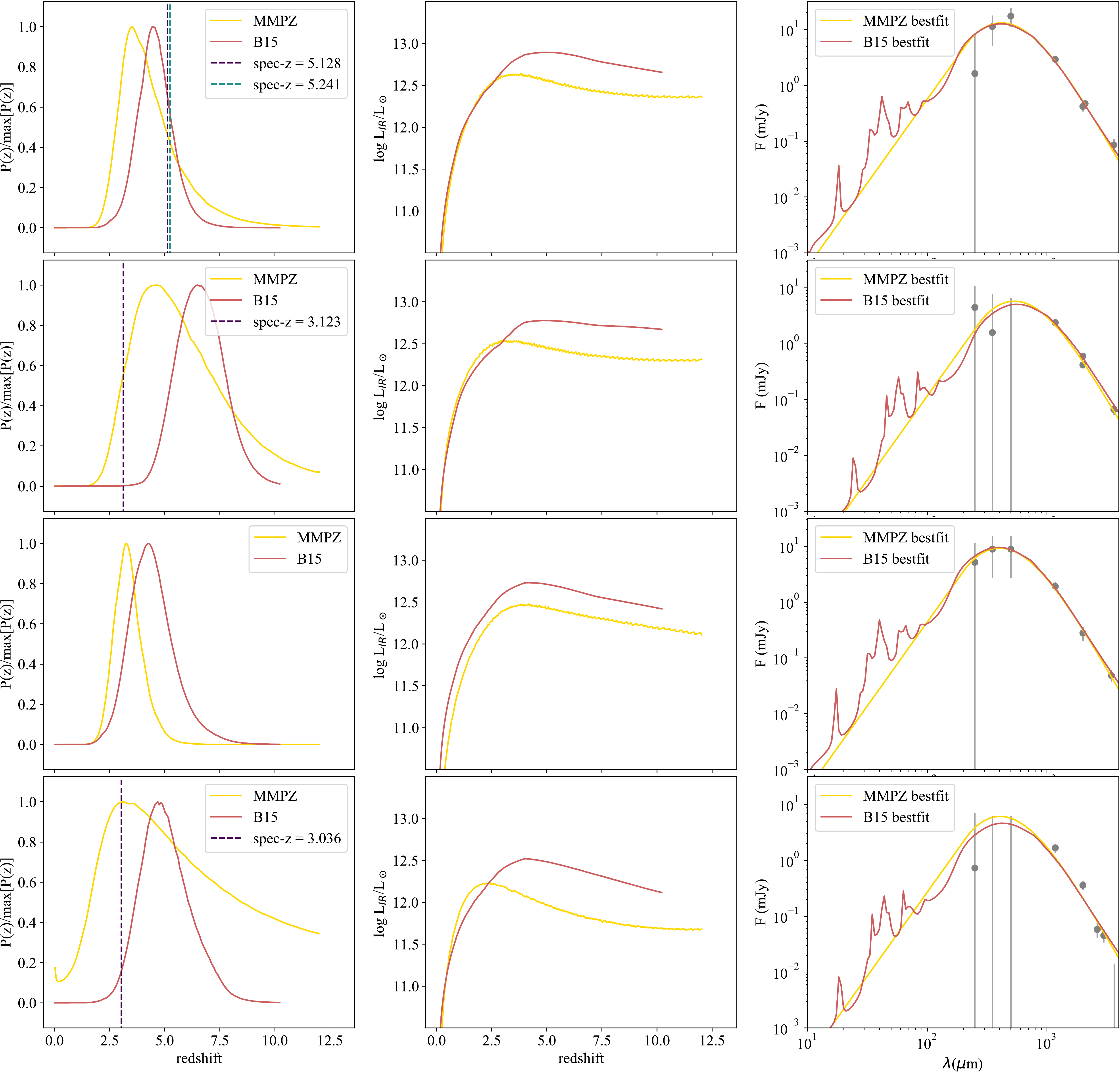}
      \caption{Results of IR template fitting on our 4 HLS sources with the \citetalias{Bethermin+15} dust templates and MMPZ method, using the SPIRE, NIKA2 and NOEMA photometry. The plots in the first column show the probability density distribution (normalized by the peak values) of each sources. The second column shows the evolution of the weighted average infrared luminosity with the redshift. The third column shows the best fit SED models with the observations. Sources from the top to the bottom are HLS-2, HLS-3, HLS-4 and HLS-22.}
      \label{hls_redshift}
   \end{figure*}
    
   From the analysis and results shown in Fig.\,\ref{hls_redshift}, we find that the redshifts from MMPZ are systematically lower than those from \citetalias{Bethermin+15} template fitting, with a typical $\Delta z/(1+z) $ of around 20\%. However, the two redshifts are still consistent within their uncertainties. The infrared luminosities returned by MMPZ are also systematically lower by $\sim$0.3\,dex, especially at redshifts beyond 3.

   The faintness and large flux uncertainties of our sources in the 3 SPIRE bands make the constraint on the peak of their IR SEDs much worse than for brighter/lensed high-z sources, which leads to large uncertainties on the estimated total IR luminosity. Compared to the template fitting with \citetalias{Bethermin+15}, MMPZ further takes the CMB heating and dimming \citep{daCunha+13} into consideration. Although this could affect the dust emissivity index $\beta$ and, as a result of $\beta$-T degeneracy, the dust temperature and IR luminosity, the $\beta$ values are all fixed to 2 in these 2 templates. Thus we consider that the inclusion of the CMB effect is not the major contributor to the differences between the results of the two template fitting methods. 
  
   The difference in the estimated total IR luminosities will be propagated to the joint photometric and spectral analysis on the source redshift in Sect.\,\ref{sc:joint_pdf_z}.

   \subsection{Joint analysis of photometric redshifts and NOEMA spectra}\label{sc:joint_pdf_z}
   
   Due to the lack of characteristic spectral features in far-IR, the photometric redshifts of our sample derived from Sect.~\ref{sc:photo-z_hls} still have large uncertainties. Searching for emission lines in the millimeter spectra provides an approach to constrain our redshifts with a significantly better accuracy. To identify the possible emission lines in the spectra, we performed a blind search in the NOEMA spectra. The NOEMA spectra are first convolved by a box kernel of 500\,km/s width, which corresponds to the typical molecular line width of bright (sub)millimeter selected galaxies \citep[e.g.,][]{Bothwell+13}. To more completely uncover the possible emission lines in these noisy spectra, we list five lines (if exist) with the highest S/N in the convolved spectra with S/N $>$ 3 in Table~\ref{tab:blind_search}. We failed to detect any lines for HLS-2 and HLS-4 with S/N $>$ 3 in our observations. For HLS-3 and HLS-22, we identify one and two detections, respectively. The "detection" at 100.628\,GHz in the HLS-22 spectrum is likely to be a glitch or noise spike with wrongly estimated uncertainty (see Appendix~\ref{sc:test_linewidth} and Fig.~\ref{plot_spike_HLS22}). With only one significant detection of an emission line, it is not possible to have an unambiguous redshift solution.\\ 
   
   \begin{table}[]
       \centering
        \caption{S/N$>$3 lines blindly detected in the NOEMA spectra.}
       \begin{tabular}{cccccc}
       \hline
       \hline
            Source & $\nu_{obs}$ & S/N \\\hline
            HLS-3 & 139.750 & 3.3 \\
            HLS-22 & 85.680 & 3.2 \\
                   & 100.628 & 3.0 \\\hline
       \end{tabular}
       \label{tab:blind_search}
   \end{table}
   
   To find the redshift solutions, we need to take additional constraints from the broad-band photometry, in particular from the total infrared luminosities at any sampled redshift in the SED fitting. From the output $\chi^2$ and IR luminosities of all models at one given redshift, we could derive the weighted average value of total infrared luminosity of the source at this redshift using Eq.\,(\ref{avg_lir_B15}):
   \begin{equation}
      L_{\mathrm{IR,avg}}(z) = \frac{\sum_{j=1}^{n} L_{\mathrm{IR,j}}(z)\times\exp{\{[\chi^2(z,j)-\sigma^2(j)]/2\}}}{\sum_{j=1}^{n}\exp{\{[\chi^2(z,j)-\sigma^2(j)]/2\}}} \,,
      \label{avg_lir_B15}
   \end{equation}

   where the $\sigma(j)$ is a weighting term to account for the deviation of the 13 model SEDs from the median of star-forming galaxies at a given redshift in \citetalias{Bethermin+15}. Indeed, at a given redshift, the \citetalias{Bethermin+15} template includes 1 median SED and 12 SEDs within $\pm$3$\sigma$ uncertainties with a spacing of 0.5$\sigma$. So when deriving the source IR luminosities at given redshifts, the $\sigma(j)$ terms should be included to account for the probability of the IR template SEDs to deviate from the median in \citetalias{Bethermin+15} model. The values of $\sigma(j)$ are thus between -3 and +3 with step of 0.5. When using the output from MMPZ, the $\sigma(j)$ will be set to 0.\\ 
   
   With a series of average IR luminosity over the redshift grid from the SED fitting, we linearly interpolate the IR luminosity at any given redshift. We further use the IR luminosity to constrain the fluxes of strong FIR-millimeter emission lines at any given redshift based on the well defined, almost redshift invariant L$_{FIR}$-L$_{line}$ relations in the form of Eq.~\ref{lir_lco}:
   \begin{equation}
      L_{line} = N \times log(L_{FIR}) + A \,.
      \label{lir_lco}  
   \end{equation}
 The luminosities and fluxes of the $^{12}$CO lines of J(1-0) to J(12-11), two transitions of [CI], and the [CII] line at 158\,$\mu$m are predicted based on various scaling relations found in the literature. The detailed information are listed in Table\,\ref{scaling_relation} and references therein. With the estimated fluxes of different line species at a given redshift, we generate a model spectrum in the frequency range of the NOEMA spectral scans and compare this model with the observations\footnote{Note that the aim of this framework is to find the redshift solution and not to precisely measure the line properties. The result of our method is not highly sensitive to different line widths and normalization of the L$_{FIR}$-L$_{line}$ scaling relation, as shown in the Appendix\,\ref{sc:test_linewidth}.}.

  \begin{table}
   \caption{Parameters of the log-linear L$_\mathrm{FIR}$-L$_\mathrm{line}$ in our analysis.} 
   \label{scaling_relation}      
   \centering 
   \begin{threeparttable}
   \begin{tabular}{c c c c c}     
   \hline\hline              
   Line Name & Rest Frequency & N & A & Scatter \\
   & [GHz] & & & dex\\  
   \hline                                   
   CO(1-0)\tnote{1} & 115.27120 & 0.99 & 1.90 & 0.30 \\ 
   CO(2-1)\tnote{1} & 230.53800 & 1.03 & 1.60 & 0.30 \\ 
   CO(3-2)\tnote{1} & 345.79599 & 0.99 & 2.10 & 0.30 \\
   CO(4-3)\tnote{2} & 461.04077 & 1.06 & 1.49 & 0.27 \\
   CO(5-4)\tnote{2} & 576.26793 & 1.07 & 1.71 & 0.22 \\
   CO(6-5)\tnote{2} & 691.47308 & 1.10 & 1.79 & 0.19 \\
   CO(7-6)\tnote{2} & 806.65181 & 1.03 & 2.62 & 0.19 \\
   CO(8-7)\tnote{2} & 921.79970 & 1.02 & 2.82 & 0.21 \\
   CO(9-8)\tnote{2} & 1036.9124 & 1.01 & 3.10 & 0.27 \\
   CO(10-9)\tnote{2} & 1151.9855 & 0.96 & 3.67 & 0.26 \\
   CO(11-10)\tnote{2} & 1267.0145 & 1.00 & 3.51 & 0.27\\
   CO(12-11)\tnote{2} & 1381.9951 & 0.99 & 3.83 & 0.28\\
   $\mathrm{[CI]}$(3P1-3P0)\tnote{3}  & 492.16065 & 1.28 & 0.00 & 0.26 \\
   $\mathrm{[CI]}$(3P2-3P1)\tnote{3} & 809.34197 & 1.28 & 0.61 & 0.50 \\
   $[\mathrm{CII]}$\tnote{4} & 1900.5369 & 1.01 & 2.84 & 0.50 \\
\hline
\end{tabular}
\begin{tablenotes}
     \item[1] \citet{Greve+14}
     \item[2] \citet{Liu+15}
     \item[3] \citet{Valentino+18}
     \item[4] \citet{DeLooze+14}
   \end{tablenotes}
     \end{threeparttable}
\end{table}

 When generating the model spectra, we assume the emission lines have Gaussian profiles with a fixed full width half maximum (FWHM) of 500\,km/s. We also linearly interpolate the L$_{FIR,med}$-z relations from the IR template fitting to a finer redshift grid to avoid missing any possible redshift solutions. 

 The spacing between adjacent redshifts in the resampled grid satisfies Eq.~\ref{eq:velspacing}, which is equivalent to a fixed spacing in velocity ($\Delta \mathrm{v}$) between adjacent redshifts: 
   \begin{equation}
    \centering
      \Delta \mathrm{z} = \frac{\Delta \mathrm{v}(1+z)}{c}\,.
      \label{eq:velspacing}
   \end{equation}
 
 We fixed the $\Delta \mathrm{v}$ to be 1/3 of the chosen FWHM, making the emission line profile to be Nyquist-sampled by the predicted line centers at the corresponding redshifts in the new grid. This ensures that emission lines in the spectra and their corresponding redshift solutions will not be missed in our analysis due to poor redshift sampling. The goodness of the model prediction at a given redshift is evaluated by log-likelihood ln($\mathcal{L}_{spec}(z)$) from the $\chi^2$ between the model spectra and the data, as given below in Eq.\,\ref{eq:lik} and Eq.\,\ref{eq:chi2_line}: 
   \begin{equation}
      \mathcal{L}_{spec}(z) \propto \exp(-\chi^2_{spec}(z)/2)\,,
      \label{eq:lik}
   \end{equation}
   \begin{equation}
      \chi^2_{spec}(z) = \sum_{\nu_i}\frac{[f_{\nu_{i,obs}}-f_{\nu_{i,model}}(z)]^2}{\sigma^2_{\nu_i}(z)}\,.
      \label{eq:chi2_line}
   \end{equation}
   
   In addition to the goodness of match between spectra and models, we further account for the goodness of SED fitting at given redshifts, $\chi^2_{SED}$(z), which is defined similarly to Eq.\,\ref{eq:chi2_line}. The joint log-likelihood at each sampled redshift reads as:
   \begin{equation}
      \mathcal{L}_{joint}(z) \propto \mathcal{L}_{spec}(z) \times \exp[-\chi^2_{SED}(z)/2].
      \label{eq:joint_likelihood}
   \end{equation}
   
   As already pointed out, we assume that the two counterparts for HLS-2 have a similar redshift and share the same FIR SED. 
   Under this assumption, the total infrared luminosity of the two NOEMA sources is thus computed based on their contributions to the total flux at 2\,mm, which are later used in deriving their final joint-likelihood of redshift.\\

   \begin{figure*}[h!]
   \centering
      \includegraphics[width=0.87\textwidth]{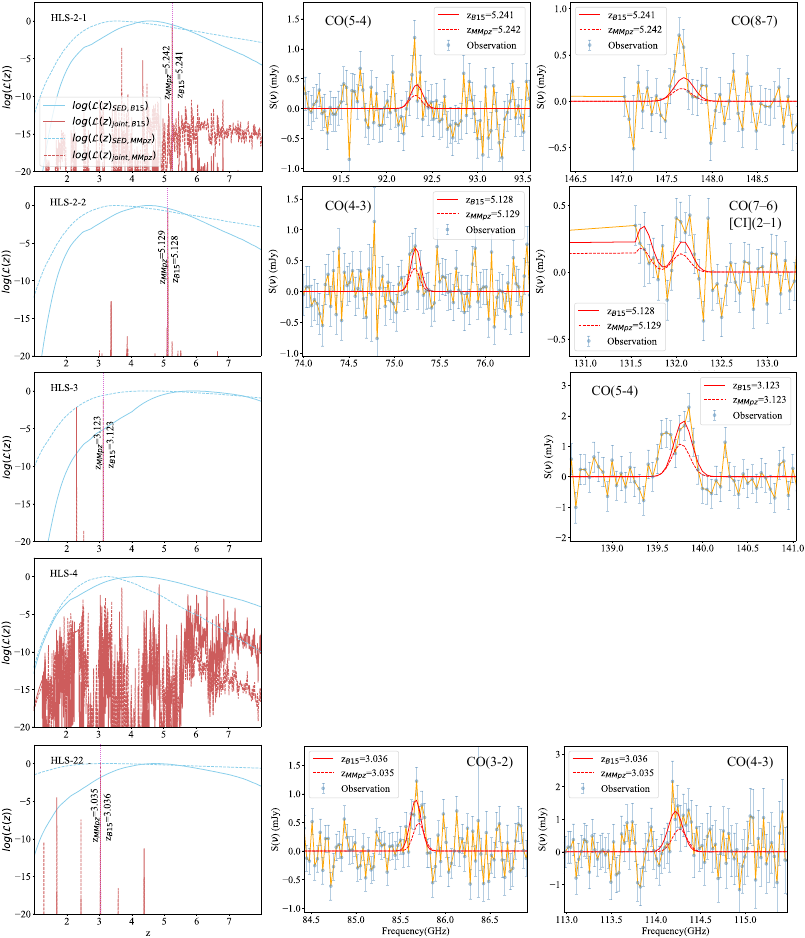}
      \caption{Joint analysis of the redshift for 4 NIKA2 HLS sources with the SED fitting outputs using the \citetalias{Bethermin+15} dust templates and MMPZ. The first row shows the normalized log-likelihood from SED fitting and the joint log-likelihood for each source after considering the information obtained from the NOEMA spectral scans. The second and third row show the cutout of the spectra around candidate spectral lines at the best redshift solutions. The lines shown in the second row are detected in the earliest band1 spectral scans (W17EL and W17FA) and those in the third row are detected in the additional follow-up observations (W18FA and S20CL). The models generated based on the fit with the \citetalias{Bethermin+15} dust templates and MMPZ at the most probable redshift are plotted as solid and dashed red lines, respectively. We emphasize again that these models are not coming from parametric fitting but are generated using an estimate on the total infrared luminosity and the L$_{CO}$(L$_{[CI]}$)-L$_{IR}$ scaling relations. Given the line profile, the spectra of HLS-3 will be analysed using a double-gaussian model. The rest of the sources will be analysed using a single-gaussian model.}
      \label{joint_redshift_HLS}
   \end{figure*}
 
   \subsection{Redshifts measurement}\label{sc:z_results}
   The results of the joint log-likelihood of redshift from photometry+spectral scan analysis on the five HLS sources are shown in Fig~\ref{joint_redshift_HLS}. For each source, we normalize the $\mathcal{L}_{spec}(z)$ to the peak value, which helps us to compare quantitatively the relative goodness of match between the model predictions and the observed spectra at different redshifts. We select all the peaks in ln($\mathcal{L}_{spec}(z)$) with an amplitude larger than -10 and a width larger than 3 samples in the redshift grid as the possible redshift solutions of our sources, using the "find\_peaks" algorithm in SciPy. Considering the large uncertainties on the total infrared luminosity of HLS sources, we further cross validate their possible redshift solutions by repeating the joint likelihood analysis using the output IR luminosity at different redshifts from MMPZ fitting, and apply the same algorithm to record the possible redshift solutions.\\
   
   Compared to the log-likelihood of redshift with photometric constraints only (see Fig.\,\ref{hls_redshift}), the joint analysis helps us to highlight significant isolated peaks for the redshift of HLS-2-1, HLS-2-2, HLS-3 and HLS-22 at z=5.241, 5.128, 3.123 and 3.036, respectively. The redshift with maximum log-likelihood value is also not sensitive to the choice of IR templates. As shown in Fig~\ref{joint_redshift_HLS}, the \citetalias{Bethermin+15} template and MMPZ find almost the same redshift where the joint log-likelihood value reaches the maximum, which further confirm their redshift solutions as listed above. For HLS-4, despite having the most accurate photometric redshift constrained through template fitting, the absence of emission line detection in the band1 spectral-scan observations results in our analysis being unable to identify significant peaks in the joint log-likelihood. Thus no reliable redshift solution is found for this source. \\
   
   For HLS-3 and HLS-22 with blindly detected candidate emission lines at S/N$>$3, our method successfully confirms the candidate lines but for that at 100\,GHz in the HLS-22 spectrum. The extremely narrow profile of the candidate detection suggests that this is likely to be a glitch, which is shown and discussed in Appendix~\ref{sc:test_linewidth}. For HLS-3, MMPZ also reveals a secondary redshift solution at z=2.299 with slightly lower log-likelihood in the analysis. This assigns the most significantly detected emission line at 139.746\,GHz to be CO(4-3), while the best solution at z=3.123 assigns it to be CO(5-4). If HLS-3 has z=3.123, we also expect to cover the CO(3-2) line in the spectral scan. Although the line is not detected at 3$\sigma$ (see Sect.~\ref{sc:properties_gas_hls}), we could not simply reject any of the two possible redshift solutions due to the the high noise level around the observed frequency. Thus, we consider the redshift of HLS-3 to be less secure than that found for the other sources with at least two lines with S/N$>$4 (HLS-2-1, HLS-2-2 and HLS-22, see Table~\ref{linemeasure}) and we will provide the estimate of HLS-3 properties based on both redshift solutions in the paper. 
   We also note that HLS-2-2 could have a secondary redshift solution at z=3.385. However, this redshift could not match with any of the two most significant emission lines found in the spectrum, which correspond to CO(4-3) and [CI](2-1) at z=5.128. Thus, we only adopt the z=5.128 solution in the following analysis.
   
   In Table~\ref{z_summary}, we summarize the redshifts from the joint analysis method (z$_{joint}$), as well as the far-IR photometric redshifts based on the two far-IR templates (z$_{B15, ms}$ and z$_{MMPZ}$). For sources with ambiguous redshift solutions, we use the two different z$_{fix}$ for the analysis. The uncertainties of z joint are conservatively given and correspond to 0.5×FWHM of the line (see Table~\ref{linemeasure}) In the following sections, the source properties are estimated at the best solution of redshift of joint analysis, or the most possible photometric redshift when no redshift solution is found in the joint analysis. These choices of redshifts of our sample are listed as z$_{fix}$ in Table~\ref{z_summary}.

\begin{table}
\caption{Summary on the joint-analyzed redshifts of NOEMA sources}           
\label{z_summary}   
\centering
\begin{tabular}{c c c c c}       
\hline\hline                 
Name & z$_{B15, ms}$ & z$_{MMPZ}$ & z$_{joint}$ & z$_{fix}$  \\    
\hline                                   
   HLS-2-1  & $4.6^{+0.8}_{-0.7}$ & $3.5^{+1.5}_{-0.8}$ & 5.241$\pm$0.003 & 5.241 \\   
   HLS-2-2  & $4.6^{+0.8}_{-0.7}$ & $3.5^{+1.5}_{-0.8}$ & 5.128$\pm$0.005 & 5.128 \\
   HLS-3  & $5.9^{+1.2}_{-0.9}$ & $4.6^{+2.5}_{-1.4}$ & 3.123$\pm$0.005 & 3.123 \\   
    &  &  & 2.299$\pm$0.004 & 2.299 \\
   HLS-4  & $4.3^{+1.0}_{-1.0}$ & $3.3^{+0.6}_{-0.6}$ & N/A & 4.3, 3.3 \\   
   HLS-22 & $4.7^{+1.3}_{-0.9}$ & $3.1^{+4.6}_{-1.4}$ & 3.036$\pm$0.003 & 3.036 \\ 
   \hline                                  
   \end{tabular}
\end{table}

   \subsection{Robustness and self-consistency of the joint analysis method}
The analysis on the source properties of our sample, as dust mass, temperature and star-formation rate, largely relies on the estimated redshifts from the joint analysis method, which is subject to assumptions on the line widths and line luminosities. To test the robustness of the redshift derived from the joint-analysis method, we make tests with model spectra of varying line widths, with NOEMA data of more limited spectral coverage and with different far-IR templates in deriving photometric redshift and predicting IR luminosity. These tests show that the redshifts of our sources from the joint analysis are reliable. Besides, we also check the self-consistency of our redshift solution by comparing our L$_{FIR}$-L$_{CO}$ correlation with the scaling relations and their scatters. These tests and discussions are presented in Appendix~\ref{sc:test_linewidth},\ref{sc:test_specoverage},\ref{sc:test_consistency} and \ref{sc:test_lfir_lco}.
   
   \section{Source properties}\label{sc:properties}
   
   \subsection{Kinematics and excitation of molecular gas of HLS sources}\label{sc:properties_gas_hls}
  We find redshift solutions to 4 NOEMA sources associated with 3 NIKA2 sources. Each of these sources has at least one emission line detected with SNR$>$4 or 2 lines with SNR at $\sim$3-4 in NOEMA spectral scans. With these redshifts, we further measure the flux and line width of the spectral lines covered by the observations, and derive the corresponding lines luminosity. We start the fitting with a single Gaussian model on the continuum-subtracted spectra of each source. No matter if they are detected with high significance, all CO/[CI] lines falling into the frequency coverage of NOEMA are considered. To make a more robust analysis on the line width, we also force the kinematics of CO and [CI] lines to be the same during the fit. For the emission line at 139.746\,GHz of HLS-3, a double Gaussian model results in an Akaike information criterion (AIC) of -17.1, comparing to the AIC of -11.3 using a single Gaussian model. This suggests an improved quality of the fit with the double-Gaussian model and thus we use this as the model for HLS-3. 
   
   The line widths, fluxes and  upper limits of CO/[CI] lines for each source are listed in Table\,\ref{linemeasure}. We measure the total flux or flux upper limit of each line by integrating the spectra within $\pm 3\sigma_{line}$ around the best-fit line center for the single-peaked lines. For HLS-3 with a double-peaked line profile (noted as the red and blue peak, respectively), we estimate the line fluxes and upper limits by integrating the spectra in the range of [f$_{center,red}$-3$\sigma_{red}$, f$_{center,blue}$+3$\sigma_{blue}$]. The corresponding CO/[CI] line luminosities or 3$\sigma$ upper limits (L'$_{line}$ and L$_{line}$) are also calculated using the following equations (from \citealt{Solomon+97}):
      \begin{equation}
      L'_{line} = 3.25\times10^7 S_{line} \Delta \mathrm{V} \nu_{obs}^{-2} D_{L}^{2}(1+z)^{-3}\, ,
      \label{L'CO}
   \end{equation}
   \begin{equation}
      L_{line} = 1.04\times10^{-3} S_{line} \Delta \mathrm{V} \nu_{rest}D_{L}^{2}/(1+z)\, ,
      \label{LCO}
   \end{equation}
where $S_{line} \Delta \mathrm{V}$ is the velocity integrated flux in Jy\,km\,s$^{-1}$ , $\nu_{rest}$ = $\nu_{obs}$(1 + z) is the rest frequency in GHz, and $D_{L}$ is the luminosity distance in Mpc.
   
   The [CI](1-0) lines of HLS-2-1 and HLS-2-2 are covered by the spectral scan but are located at the noisiest edges of the NOEMA sidebands. This makes their upper limits of little scientific value and thus we discard them from the table. The CO(7-6) line of HLS-2-2 is marginally detected but only partly covered by our observations. For this line, we use the output parameters from the spectral line fitting to constrain its flux using a complete Gaussian profile.

   \begin{table*}
   \caption{Emission line fluxes, luminosities and widths of 4 sources in the HLS field.}          \label{linemeasure}      
   \centering                                      
   \begin{tabular}{c c c c c c c}          
   \hline\hline                        
   Source & Line & F$_{obs}$ & S$_{line}$ & L'$_{line}$ & L$_{line}$ & FWHM \\
          &      & (GHz) & (mJy km/s) & 10$^9$ K km/s pc$^2$ & 10$^7$ L$_\odot$ & km/s \\
   \hline                                   
   HLS-2-1 & CO(4-3) & 73.852 & $<$240 & $<$15.0 & $<$4.7 & 254 \\ 
              & CO(5-4) & 92.309 & 246$\pm$65 & 9.8$\pm$2.6 & 6.0$\pm$1.6 & \\
              & CO(8-7) & 147.658 & 193$\pm$44 & 3.0$\pm$0.7 & 7.6$\pm$1.7 & \\
   HLS-2-2 & CO(4-3) & 75.226 & 309$\pm$96 & 18.6$\pm$5.8 & 5.8$\pm$1.8 & 487 \\
              & CO(5-4) & 94.027 & $<$513\tablefootmark{1} & $<$11.2\tablefootmark{1} & $<$6.9\tablefootmark{1} &  \\
              & CO(7-6) & 131.618 & $<$164\tablefootmark{1} & $<$3.2\tablefootmark{1} & $<$5.4\tablefootmark{1} &  \\
              & [CI](2-1) & 132.057 & 196$\pm$50 & 3.8$\pm$1.0 & 6.5$\pm$1.7 &  \\
              & CO(8-7) & 150.406 & $<$151\tablefootmark{1} & $<$2.3\tablefootmark{1} & $<$5.7\tablefootmark{1} &  \\
   HLS-3 & CO(3-2)\tablefootmark{2} & 83.856 & $<$665\tablefootmark{1} & $<$32.8\tablefootmark{1} & $<$4.3\tablefootmark{1} & 752 \\
              & CO(5-4)\tablefootmark{2} & 139.746 & 1347$\pm$164 & 23.9$\pm$2.9 & 14.7$\pm$1.8 &  \\
         & CO(4-3)\tablefootmark{3} & 139.746 & 1347$\pm$164 & 22.3$\pm$2.8 & 7.0$\pm$0.9 & \\
   HLS-22  & CO(3-2) & 85.691 & 521$\pm$82 & 24.5$\pm$3.8 & 3.2$\pm$0.5 & 508 \\
              & CO(4-3) & 114.250 & 719$\pm$170 & 19.0$\pm$4.5 & 6.0$\pm$1.4 &  \\
   \hline                                             
   \end{tabular}
   \tablefoot{\tablefoottext{1}{Upper limits are given at 3$\sigma$.} \tablefoottext{2}{Considering z=3.123.} \tablefoottext{3}{Considering z=2.299. Given the line detected at 139.746\,GHz, no other CO/[CI] line is expected to fall within the spectral coverage.}}
   \end{table*}

\begin{figure*}
   \centering
      \includegraphics[clip=true,width=1.0\textwidth]{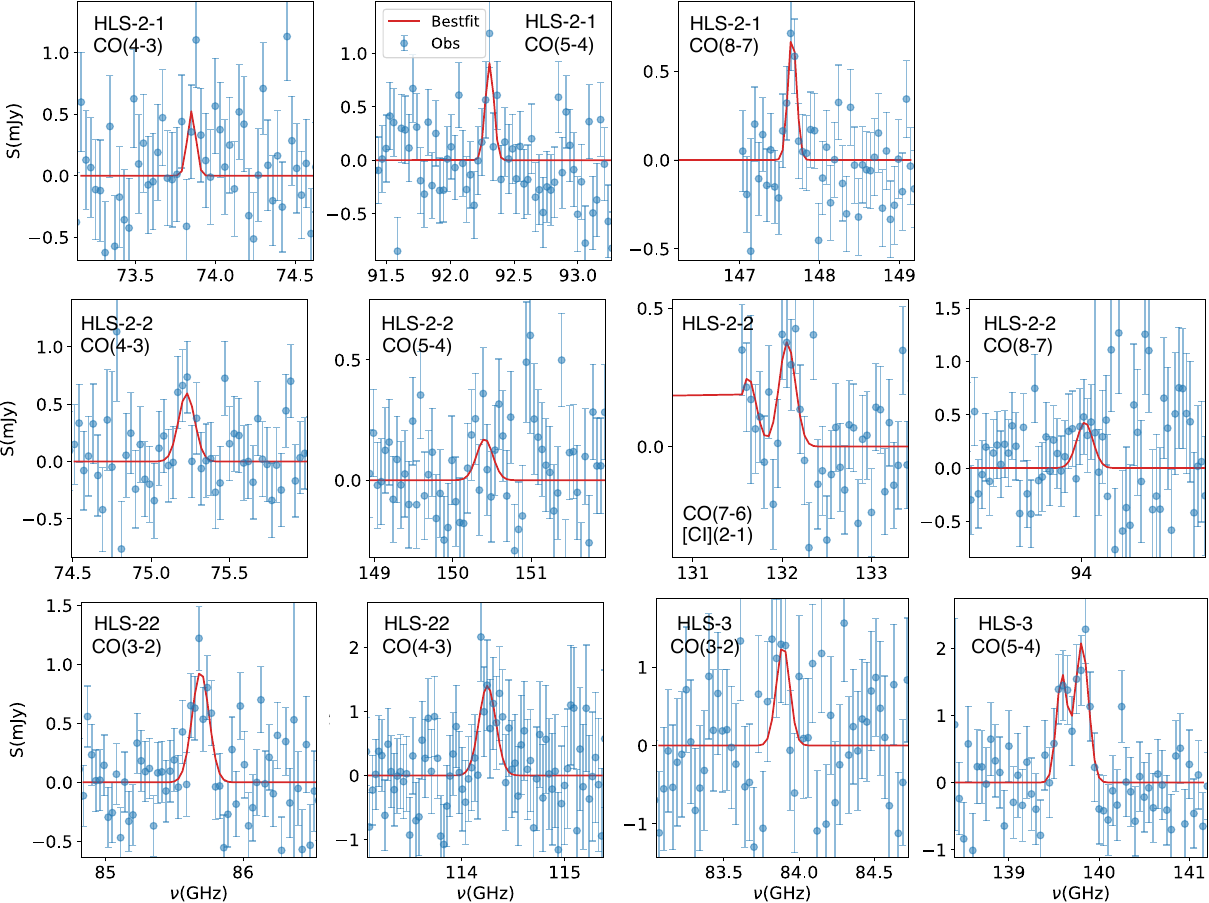}
      \caption{Observations and best fits on spectral lines, including both detections and upper limits. The best-fit model of each line is shown with the red solid line. For HLS-3, we show the two lines assuming z=3.123. The luminosity of the CO(3-2) line has a best-fit value consistent with zero.}
      \label{linefit}
   \end{figure*}

   Figure\,\ref{linefit} shows the best-fit models for each CO/[CI] line. The line identification in each panel are presented assuming the best redshift solution of each source, as listed in Table~\ref{z_summary}. The spectra have the same channel width as the ones we used for the joint analysis. From the best-fit parameters we find the line widths are generally consistent with the assumption we made during the redshift search in Sect.\,\ref{sc:joint_pdf_z}, with an average FWHM of 500\,km/s. Although previous observations reveal that the integrated [CI] and CO lines from the same high-z galaxies may have different line widths and line profiles \citep{Banerji+18}, fixing or relaxing the velocities and widths of the different lines during our analysis does not significantly change the quality of the best-fit model. 
   
   The observations of the 2 sources associated with HLS-2 cover both mid-J (CO(4-3) or CO(5-4)) and high-J CO lines (CO(7-6) or CO(8-7)), allowing us to roughly estimate the conditions of their molecular gas and compare with other DSFGs at similar redshifts using luminosity ratios (expressed in K\,km/s\,pc$^2$). For HLS-2-1, the L'$_{CO(8-7)}$/L'$_{CO(5-4)}$ is 0.31$\pm$0.11, which is consistent with the values found in typical high-z SMGs with low excitation \citep{Bothwell+13} but lower than the reported value of some starburst galaxies and luminous quasars at similar redshift \citep{Rawle+14,Li+20}. The L'$_{CO(7-6)}$/L'$_{CO(4-3)}$ and CO(8-7)/CO(4-3) of HLS-2-2 are $<$0.17 and $<$0.12, respectively. These values are even lower than the typical value of high-z SMGs but still consistent with the low excitation ISM found in the "Cosmic Eyelash" \citep{Danielson+11}. \\
   
   Apart from CO detections/upper limits, the detection of [CI](2-1) for HLS-2-2 also provides an additional insight on the state and condition of its molecular gas reservoir. Previous studies suggested that L'$_{CO(7-6)}$/L'$_{[CI](2-1)}$ could be used to distinguish secular-evolved (low values) and merger driven (high values) systems. The [CI] dominated systems with L'$_{CO(7-6)}$/L'$_{[CI](2-1)}$ around or below 1 are generally found for secular-evolved disk dominated galaxies \citep{Andreani+18}. As neutral carbon could be more easily excited, the low values in secular-evolved systems indicate lower gas excitation/more abundant low density gas.
   The low L'$_{CO(7-6)}$/L'$_{[CI](2-1)}$ suggests a low excitation molecular gas reservoir in HLS-2-2 and is consistent with the low L'$_{CO(7-6)}$/L'$_{CO(4-3)}$ and L'$_{CO(8-7)}$/L'$_{CO(4-3)}$ measured in the same galaxy.
   For HLS-2-1, we estimate L'$_{CO(7-6)}$/L'$_{[CI](2-1)} <$0.9, which is in agreement with the values found in secularly-evolved galaxies.\\
   
   For the rest 2 sources with CO detection of less separated quantum numbers J, our observations find L'$_{CO(5-4)}$/L'$_{CO(3-2)}>$0.73 in HLS-3 and L'$_{CO(4-3)}$/L'$_{CO(3-2)}$=0.78$\pm$0.22 in HLS-22, respectively. The value for HLS-22 is generally consistent with the average CO SLED of high-z SMGs in \citet{Bothwell+13}, being similar to the case of HLS-2-1. On the contrary, HLS-3 has a L'$_{CO(5-4)}$/L'$_{CO(3-2)}$ ratio higher than typical SMGs in \citet{Bothwell+13} and resembles the average of the SPT sample \citep{Spilker+14} or the local starburst galaxy M82 \citep{Carilli+13} with higher excitation. However, the observations on these 2 sources do not cover higher-J CO lines like for HLS-2, which traces warmer and denser components in the molecular gas reservoir. Thus, with these 2 line ratio measurements, it is more difficult to conclude.
  
   \subsection{Dust mass and dust temperature}\label{sc:properties_dust_mbb}
   The Far-IR continuum emission of star-forming galaxies could be well represented by a single temperature modified black-body model from which the dust temperature (T$_{dust}$), dust emissivity index ($\beta$) and total dust mass (M$_{dust}$) can be derived. At high redshift, the increasing temperature of the cosmic microwave background (CMB) reduces the contrast of star-forming galaxy emissions in the (sub-)millimeter and changes the apparent shape of the spectrum at these frequencies. Considering the impact of the CMB, the observed modified black-body emission of high-z SMG could be expressed as Eq.\,\ref{mbb_thin_cmb} using the optical-thin assumption \citep{daCunha+13} :
   \begin{equation}
      S(\nu) = \frac{1+z}{d_L^{2}}M_{dust}\kappa(\nu)\left[B\left(\nu, \frac{T_{dust}}{(1+z)}\right)-B\left(\nu,\frac{T_{CMB,z}}{(1+z)}\right)\right]\,.
      \label{mbb_thin_cmb}
   \end{equation} 
   
  The dust emissivity $\kappa$($\nu$) in far-IR can be described by a single power law:
   \begin{equation}
      \kappa(\nu) = k_0 \left( \frac{\nu}{\nu_0} \right) ^{\beta}\,,
   \end{equation} 
 where $k_0$ stands for the absorption cross section per unit dust mass at a given specific frequency $\nu_0$. Here we take $k_{0, 850\mu m}$=0.047\,m$^2$/kg from \citet{Draine+14} (see also \citealt{Berta+21}).
   
\begin{figure*}
   \centering
      \subfloat[]{\includegraphics[width=8cm]{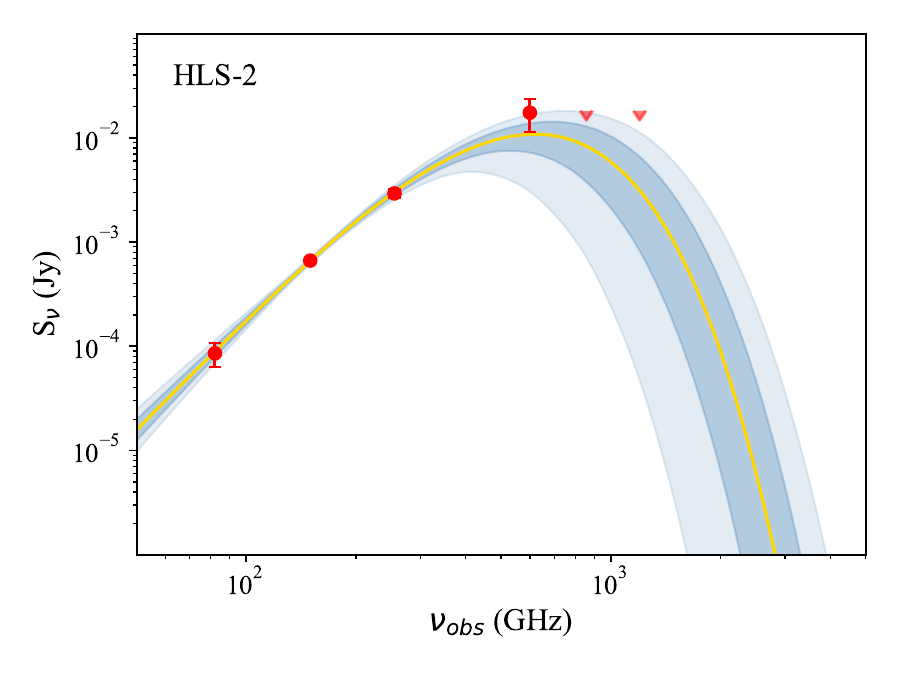} }
      \qquad
      \subfloat[]{\includegraphics[width=7.75cm]{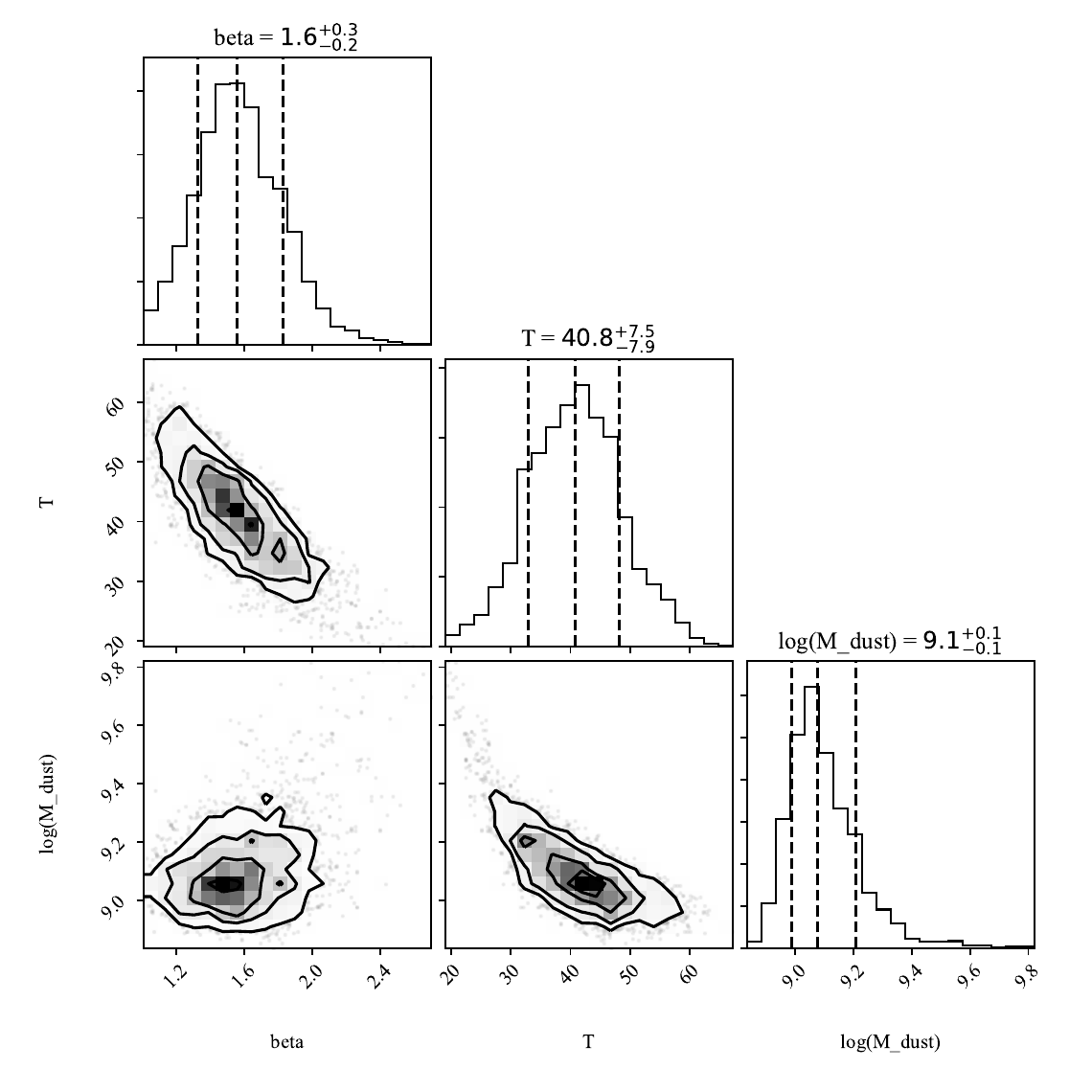}}
      \caption{Example of a modified black-body fitting on the far-IR to millimeter photometric data for one of our galaxy. (a) Photometric data and best-fit model. $\pm$1$\sigma$ and $\pm$2$\sigma$ uncertainties of the model are shown with blue shades of different transparency. (b) Corner plot of the posterior distribution of the 3 parameters. The contours correspond to 1, 1.5 and 2\,$\sigma$ in the 2D histogram.}
      \label{mbbfit_explot}
   \end{figure*}

   We perform MCMC fitting (using the PyMC3 package) on our far-IR to millimeter photometric data using the model given in Eq.~\ref{mbb_thin_cmb}. The two sources associated with HLS-2 are fitted using their integrated flux, as they have very similar redshifts and their individual fluxes at the far-IR could not be obtained with the low resolution SPIRE data. 
    We adopt uniform priors for T$_{dust}$ and M$_{dust}$ and a flat prior between 1 and 3 for dust emissivity $\beta$. We constrain the temperature to be between $T_{CMB}$ at the given redshifts and 80\,K. The redshift values are fixed to z$_{fix}$ given in Table~\ref{z_summary}. Figure~\ref{mbbfit_explot} shows as an example the best-fit modified black-body model for HLS-2, as well as the 1$\sigma$ and 2$\sigma$ confidence intervals. For HLS-22, the original fit with free parameters lead to a non-physically low dust temperature at 16\,K and a high $\beta$ larger than 3, which is due to the poor observational constraints at 3\,mm and $<$500 $\mu$m. Thus, we perform a constrained modified black-body fit with a fixed $\beta$ of 1.8, consistent with the average of the other HLS sources. 
 We list the estimated dust temperature, mass and emissivity index in Table\,\ref{MBB_dust_properties}. \\

\begin{table*}
\caption{Dust properties of the HLS sources from optical-thin modified black-body fitting.} 
\label{MBB_dust_properties}      
\centering                       
\begin{tabular}{c c c c c c c}   
\hline\hline                     
Source & z$_{fix}$ & log(M$_{dust, MBB}$/M$_\odot$) & T$_{dust, MBB}$ & $\beta$ & L$_{FIR}$(50-300$\mu$m) & SFR \\
       &   &   & K &    &  10$^{12}$L$_{\odot}$ & M$_\odot$/yr\\
\hline                        
HLS-2 & 5.2 & 9.1$^{+0.1}_{-0.1}$ & 41$^{+8}_{-8}$ & 1.5$^{+0.3}_{-0.3}$ &  5.0$^{+1.6}_{-1.5}$ & 9.6$^{+3.0}_{-3.1}\times$10$^2$ \\ 
HLS-3 & 3.1 & 9.5$^{+0.3}_{-0.2}$ & 23$^{+7}_{-6}$ & 1.8$^{+0.4}_{-0.3}$ &  1.0$^{+0.8}_{-0.5}$ & 1.9$^{+1.6}_{-0.9}\times$10$^2$\\
 & 2.3 & 9.7$^{+0.3}_{-0.2}$ & 18$^{+6}_{-5}$ & 1.8$^{+0.4}_{-0.4}$ &  0.5$^{+0.4}_{-0.2}$ & 0.9$^{+0.9}_{-0.5}\times$10$^2$\\
HLS-4 & 4.3 & 8.9$^{+0.3}_{-0.2}$ & 34$^{+11}_{-10}$ & 1.8$^{+0.4}_{-0.4}$ & 2.4$^{+1.5}_{-1.2}$ & 4.6$^{+2.8}_{-2.3}\times$10$^2$ \\
 & 3.3 & 9.2$^{+0.3}_{-0.2}$ & 28$^{+8}_{-9}$ & 1.8$^{+0.4}_{-0.3}$ & 1.5$^{+1.0}_{-0.9}$ & 2.9$^{+2.0}_{-1.7}\times$10$^2$ \\
HLS-22 & 3.0 & 8.9$^{+0.1}_{-0.1}$ & 31$^{+4}_{-5}$ & 1.8 & 1.4$^{+0.8}_{-0.6}$ & 2.8$^{+1.6}_{-1.3}\times$10$^2$ \\
\hline                                             
\end{tabular}
\end{table*}

   We derive a dust mass of $\sim$ 10$^9$\,M$_\odot$, which is consistent with the dust masses derived for bright SMGs selected from blind single-dish surveys \citep{Santini+10,Miettinen+17}. The abundant dust indicates that these high-z dusty star-forming galaxies have already experienced a rapid metal enrichment in the first few billion years of the Universe. The dust emissivity $\beta$ of our sample (excluding HLS-22) has a median value of 1.75, which is also consistent with the values found in a variety of galaxies across the cosmic time. \\
   
   We also measure the far-IR luminosities (L$_{FIR}$) by integrating the model SEDs between 50 and 300\,$\mu$m at the rest-frame of each source. The L$_{FIR}$ are listed in Table~\ref{MBB_dust_properties}. Our observations could not properly constrain and model the mid-IR emission of galaxies. Thus, we extrapolate our L$_{FIR}$ (50-300\,$\mu$m) to the total infrared luminosity (L$_{IR}$, 3-1000\,$\mu$m) by multiplying L$_{FIR}$ by a factor of 1.3, based on the calibrations given in \citet{Gracia-Carpio+08}.  We further derive the star formation rates, SFR, based on the standard scaling relations from \citep{Kennicutt+12}. The corresponding results are also listed in Table~\ref{MBB_dust_properties}. \\
   
\begin{figure*}[tb]
   \sidecaption
      \includegraphics[clip=true,width=.50\textwidth]{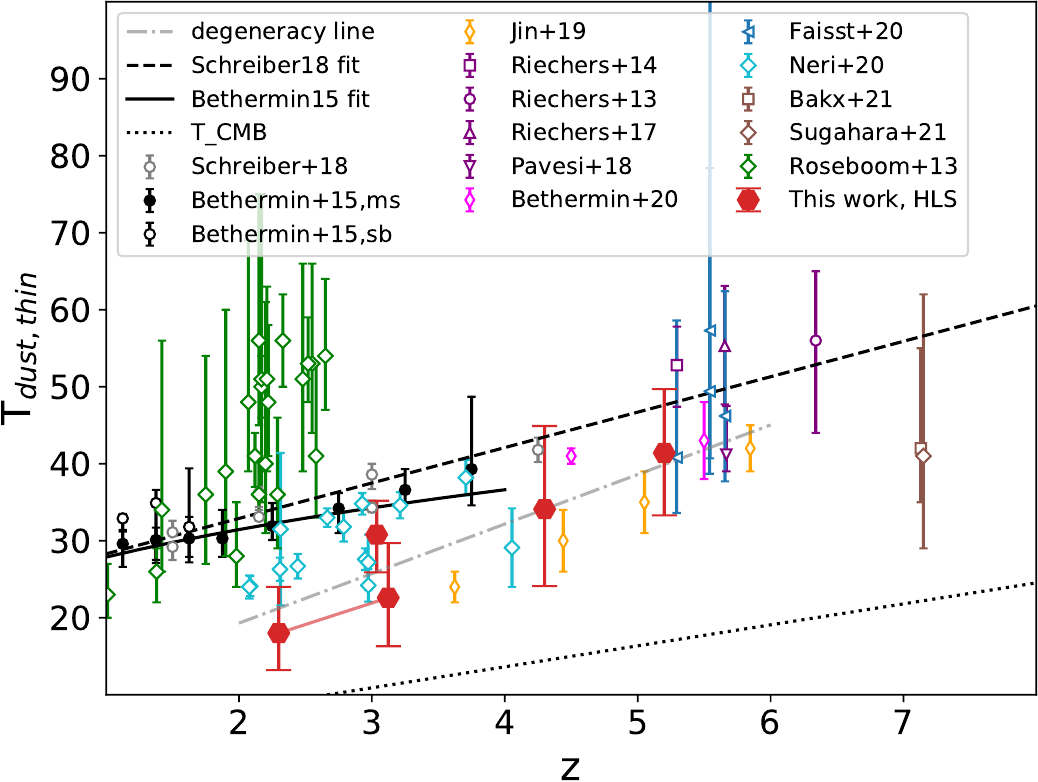}
      \caption{Dust temperature versus the redshift of high-z DSFGs in our sample, and DSFGs/LBGs from literature. The dust temperatures of these sources are all measured with the optically-thin modified black-body model. The corresponding references are listed in the legend. We also overlaid the average T$_{dust}$-z relation of main sequence galaxies derived by \citet{Schreiber+18} and \citet{Bethermin+15} based on observational data. For HLS-4 with only photometric redshift, we show with the dash-dotted grey line the degeneracy between dust temperature and redshift. The T$_\mathrm{dust}$ of HLS-3 at the two possible redshifts are both plotted and connected by the red solid line.}
      \label{tdust_z}
   \end{figure*}

   The dust temperature of our sample varies from 18 to 41\,K. Fig.~\ref{tdust_z} shows the comparison between the dust temperature of our sample with other DSFGs and star-forming galaxies \citep{roseboom+13,Riechers+14,Riechers+17,Pavesi+18,Bethermin+20,Faisst+20,Neri+20,Bakx+21,Sugahara+21} at different redshifts. Similar to our analysis, the literature dust temperatures here for comparison are 
 all derived under the optically-thin assumption. The average dust temperature of normal star-forming and starburst galaxies from \cite{Bethermin+15} and \cite{Schreiber+18} are also shown as baselines of comparison at different redshifts. 
 We find large scatters in the dust temperature of our sample with respect to these empirical T$_{dust}$-(z) scaling relations on star-forming/starburst galaxies. Among our HLS and literature sample, HLS-3 shows one of the lowest dust temperatures of 23(18)\,K, while the other three sources at higher redshifts have higher dust temperature not distinctive to literature DSFGs and the average dust temperature of normal star-forming galaxies. DSFGs with apparently cold temperatures have been reported by some studies in recent years \citep{Jin+19,Neri+20}. Similarly to these galaxies, the redder/colder far-IR SED of HLS-3 could resemble normal DSFGs at much higher redshift in SED modeling, which explains the significant deviation of its far-IR photometric redshift from its spectroscopic redshift based on the \citet{Bethermin+15} SED template. At fixed L$_{IR}$, we expect galaxies with colder dust temperature to be brighter at 1.2\,mm and thus these galaxies are more favored by the selection of candidate high-z DSFGs based on red far-IR to millimeter colors.
   
   \subsection{Molecular gas mass}\label{sc:properties_gasmass_hls}
   Both the CO emission lines and dust continuum in the Rayleigh-Jeans tail of the far-IR SED have been widely used to estimate the amount of molecular gas in galaxies \citep{Carilli+13,Hodge+20}. In this section we measure the molecular gas mass for our sample and cross-validate the results with various methods. \\
   
   The detections/constraints on CO emission lines enable the estimate of the molecular gas mass using the CO luminosity to molecular gas mass conversion factor $\alpha_{CO}$. Robust estimations of the conversion factor are mostly made on the lowest J transition, CO(1-0), while CO detections in our sample start from CO(3-2) to CO(5-4). Thus, we need to convert the luminosities of the lowest-J CO line detected in our observations to CO(1-0), in addition to the assumptions on the $\alpha_{CO}$ conversion factor between CO(1-0) luminosity to molecular gas mass. In our case, we take the advantage of multiple line detections/flux upper limits in the NOEMA spectra of each source to find the matched cases in the literature and roughly estimate the CO(1-0) luminosity and molecular gas mass. As described in Sect.~\ref{sc:properties_gas_hls}, for each source, we compare their CO luminosity ratios with literature results and find the cases with CO SLEDs that could reproduce the observed values. For HLS-2-1 and HLS-22, we apply L'$_{CO(5-4)}$/L'$_{CO(1-0)}$=0.32 and L'$_{CO(3-2)}$/L'$_{CO(1-0)}$=0.52 from the average SLED of unlensed SMGs in \citet{Bothwell+13}. For HLS-2-2 and HLS-3, we find the SLEDs of the "Cosmic Eyelash" \citep{Danielson+11} and the average of SPT SMGs \citep{Spilker+14} could well reproduce the observed luminosity ratios. We scaled the luminosities of the lowest J CO lines observed in these two sources to their CO(1-0) luminosities by assuming L'$_{CO(4-3)}$/L'$_{CO(1-0)}$=0.50 and L'$_{CO(5-4)}$/L'$_{CO(1-0)}$=0.72.
   
   Having the L'$_{CO(1-0)}$ (see Table.~\ref{mgas_sources}), we then estimate the total molecular gas mass with a fixed conversion factor $\alpha_{CO}$. As the derived star formation rates do not reveal solid evidence of on-going starburst in our sample, we adopt the typical Milky Way value of $\alpha_{CO}$=4.36M$_\odot$(K km/s pc$^2$)$^{-1}$. The estimated molecular gas are also listed in Table.~\ref{mgas_sources}.
   
\begin{table*}
\caption{Molecular gas properties of HLS sources.}
\label{mgas_sources}      
\centering                                      
\begin{tabular}{c c c c c c c}          
\hline\hline                        
Source & z$_{fix}$ & L'$_{CO(1-0),scaled}$ & L$_{850\mu m}$ & M$_{gas,CO}$ & M$_{gas,S16}$ & $\tau_{dep}$ \\
       &  & 10$^9$ K km/s pc$^2$ & 10$^{31}$ erg/s/Hz & 10$^{10}$M$_\odot$$\times$($\alpha_\mathrm{CO}/4.36$) & 10$^{10}$M$_\odot$$\times$($\alpha_\mathrm{CO}/6.5$) & Myr$\times$($\alpha_\mathrm{CO}/6.5$) \\
\hline                                   
HLS-2-1 & 5.241 & 30.7$\pm$8.1 & 0.90$^{+0.18}_{-0.14}$ & 13.4$\pm$3.5 & 10.6$^{+2.1}_{-1.7}$ & 2.2$^{+1.0}_{-0.6}\times$10$^2$ \\ 
HLS-2-2 & 5.128 & 37.2$\pm$11.5 & 0.87$^{+0.18}_{-0.14}$ & 16.2$\pm$5.0 & 10.9$^{+2.1}_{-1.7}$ & 2.2$^{+1.0}_{-0.6}\times$10$^2$ \\ 
HLS-3   & 3.123 & 35.5$\pm$4.3 & 2.26$^{+0.56}_{-0.40}$ & 15.5$\pm$1.9 & 26.9$^{+6.7}_{-4.8}$ & 1.9$^{+1.8}_{-0.9}\times$10$^3$ \\ 
    & 2.299 & 36.9$\pm$4.6 & 2.55$^{+0.69}_{-0.36}$ & 16.1$\pm$2.0 & 30.4$^{+8.3}_{-4.3}$ & 3.2$^{+5.1}_{-1.7}\times$10$^3$ \\ 
HLS-4   & 4.3 & ---- & 1.00$^{+0.28}_{-0.25}$ & ---- & 11.9$^{+3.3}_{-3.0}$ & 2.0$^{+1.9}_{-0.8}\times$10$^2$ \\ 
        & 3.3 & ---- & 1.33$^{+0.30}_{-0.28}$ & ---- & 15.9$^{+3.6}_{-3.3}$ & 4.2$^{+4.4}_{-1.6}\times$10$^2$ \\
HLS-22  & 3.036 & 47.06$\pm$7.38 & 0.90$^{+0.18}_{-0.14}$ & 20.5$\pm$3.2 & 10.7$^{+2.1}_{-1.7}$ & 7.6$^{+10.0}_{-3.7}\times$10$^2$ \\ 
\hline                                             
\end{tabular}
\end{table*}

For all sources, we also provide an estimate on their molecular gas mass using their continuum emission in the Rayleigh-Jeans tail. Following the calibration described in \citet{Scoville+16}, we derive the luminosity of dust emission at rest-frame 850$\mu$m. We use the series of optically-thin modified black-body models generated by the combinations of parameters explored by the MCMC fit to interpolate/extrapolate the dust luminosity at rest-frame 850$\mu$m, L$_{850\mu m}$. Using the conversion factor for SMGs from \citet{Scoville+16} (L$_{850\mu m}$/M$_{gas}$ = 8.4$\times$10$^{19}$ erg/s/Hz/M$_\odot$), we derive the molecular gas mass from the Rayleigh-Jeans dust emission, noted as M$_{gas,S16}$, and shown in Table\,\ref{mgas_sources}. We find the differences between the molecular gas masses estimated by the two methods are within a factor of 2. Our galaxies have molecular gas mass of 1-3$\times$10$^{11}$\,M$_\odot$, suggesting a gas-rich nature. For HLS-4, we derive a similarly massive molecular gas reservoir as that of the 4 sources with CO detections. \\

One of the primary source the uncertainty of molecular gas mass measurement comes from the CO-to-H2 conversion factor ($\alpha_\mathrm{CO}$). We applied a typical Milky-way $\alpha_\mathrm{CO}$ value of 4.36 to the CO(1-0) luminosities. The alternative estimate, based on \citep{Scoville+16}, has a equivalent $\alpha_\mathrm{CO}$ of 6.5. However, previous studies find starburst galaxies could have much lower $\alpha_\mathrm{CO}$ compared to the Milky Way like values typical for normal star-forming galaxies \citep[e.g][]{Downes+98,Tacconi+08}. The exact value of $\alpha_\mathrm{CO}$ at high redshift is still highly uncertain. Although there are evidence for $\alpha_\mathrm{CO}$ as large as the Milky Way in high-z SMGs, starburst-like values are also prevalently used in previous studies. This would introduce differences of a factor of 5-7 in molecular gas mass measurements. The impact of $\alpha_\mathrm{CO}$ is accounted for in the values of molecular gas mass and gas depletion time given in Table~\ref{mgas_sources}.

Combining the measurements from Tables\,\ref{MBB_dust_properties} and \,\ref{mgas_sources}, we derive the gas-to-dust ratios of the 4 sources (HLS-2-1, HLS-2-2, HLS-3 and HLS-22) with relatively secure spectroscopic redshifts and dust-independent molecular gas mass measurements from CO lines. With the assumption of a Milky Way like $\alpha_{CO}$, our analysis yields an average gas-to-dust ratio of 113, which is in line with values found in local and high-z massive galaxies \citep{Santini+10,Remy-Ruyer+14,DeVis+19,Rujopakarn+19} and consistent with the values expected at solar metallicity \citep{Leroy+11,Magdis+12,Shapley+20}. 
A lower, starburst-like $\alpha_{CO}$ will lead to an average gas-to-dust ratio 5 to 8 times lower, which is also consistent with the results of \citet{Rowlands+14} under similar assumptions but still at the extreme values. Such abundant dust in ISM could be difficult to explain unless the sources are already enriched to super-solar metallicity at $z=3-5$ \citep{Chen+13,Santini+14} or (and) they are undergoing vigorous merger+starburst events \citep{Silverman+18}. \\

\begin{figure*}[tb]
 \sidecaption
      \includegraphics[clip=true,width=0.5\textwidth]{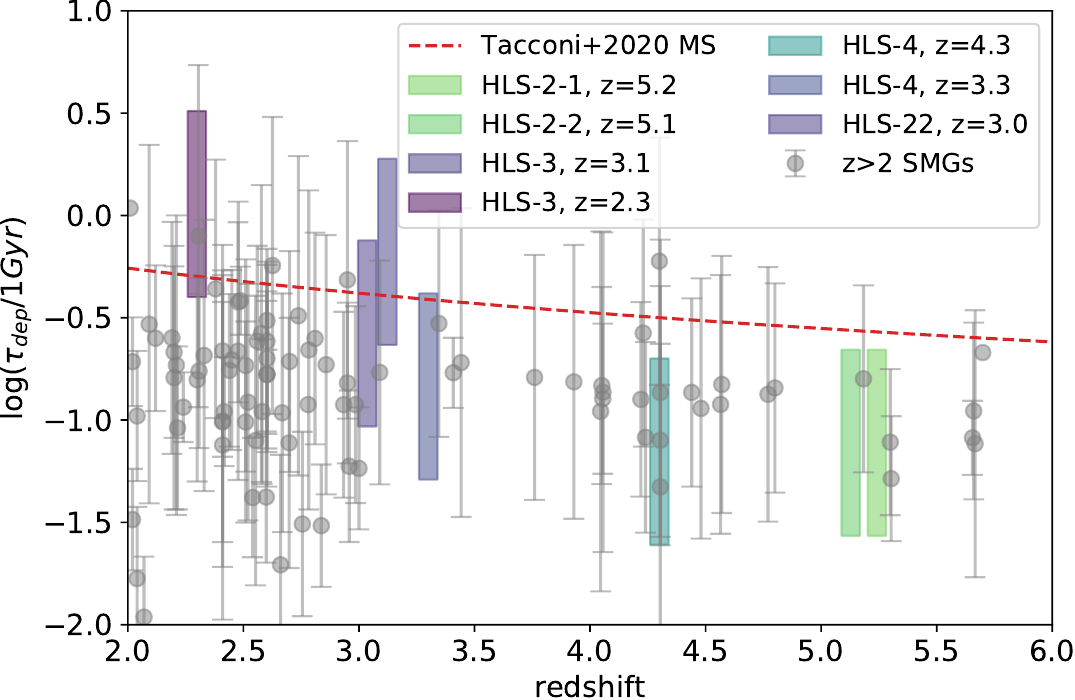}
      \caption{The gas depletion time of HLS sample based on the molecular mass from Rayleigh-Jeans dust emission and the star formation rate from far-IR luminosities. The red dashed line shows the redshift evolution of gas depletion time of main sequence galaxies from \citet{Tacconi+20}. The grey dots show the gas depletion time of z$>2$ SMGs based on the data summarized in \citet{Dunne+22}.}
      \label{gas_dep}
   \end{figure*}

We finally derive the depletion time of the molecular gas in each galaxy using the molecular gas mass and the SFR. For the molecular gas mass, we use M$_{gas,S16}$ to keep the measurement consistent among all galaxies with or without CO line detections. The results are also given in the last column of Table~\ref{mgas_sources}.  Fig~\ref{gas_dep} shows the gas depletion time of HLS sources compared to high-z main sequence galaxies \citet{Tacconi+20} and SMGs \citep{Dunne+22}. Considering the uncertainties of $\alpha_{CO}$, the plot marks the $\tau_{dep}$ in rectangles, with the upper and lower bounds at the $\tau_{dep}$ derived using $\alpha_{CO}$ values of 6.5 and 0.8, respectively. 
Fig.~\ref{gas_dep} shows that most of HLS sources have short gas depletion time of a few hundred Myrs, which is typical among high-z SMGs in \citep{Dunne+22}. The only exception, HLS-3, shows possibly long gas depletion time up to a few Gyrs and being comparable to main sequence galaxies at the same redshift. Remarkably, HLS-3 also has the largest millimeter continuum size (2.8"$\times$0.7", see Table~\ref{cont_noema_pos_shape}) and the lowest dust temperature ($23^{+7}_{-6}$K at z=3.123 or $18^{+6}_{-5}$K at z=2.299, see Table~\ref{MBB_dust_properties}). These atypical properties among SMGs, in addition to its long gas depletion time, suggest that HLS-3 is more likely a massive main sequence galaxy under secular evolution. As reported in Table~\ref{cont_noema_pos_shape}, HLS-2-1 and HLS-3, are already partially resolved in dust continuum with compact NOEMA configurations. This could further suggests extended distributions of the molecular gas reservoir/disk.

\section{A possible over-density of DSFGs at z=5.2 \label{sc:overdensity}}

It is found that HLS 2-1 and HLS 2-2, separated by 12\,arcsec on the map, have both a redshift of $\sim$5.2. They are also located within 2\,arcmin from \hls\, a bright lensed DSFG firstly found by the Herschel Lensing Survey at a similar redshift of z=5.243\citep{Egami+10,Combes+12,Rawle+14}. At this redshift, their projected separation in the sky corresponds to a physical transverse distance of $\sim$800\,kiloparsec. 

Given the close spectroscopic redshifts of HLS-2-1 and \hls\, the physical distance between these two sources is given by their transverse distance D$_t$=796\,kpc, computed following $D_t = D_A \times \theta_{sep}$.
The physical distance between HLS-2-1 (or \hls\,) to HLS-2-2, which are separated in both redshift and sky coordinates, is approximately estimated using the following equations:
\begin{equation}
    D_{los} = [D_c(z1)-D_c(z2)]/(1+\bar{z})\,, \\
    D = \sqrt{{D_t}^2+{D_{los}}^2}\,.
\end{equation}
We derive physical distances (D) of $\sim$9.4\,Mpc between HLS-2-1/\hls\, to HLS-2-2 \footnote{The bias of peculiar radial velocity estimate might not be a large effect for HLS-2-2 to the other 2 sources. For HLS-2-1 to \hls, if we consider a peculiar radial velocity up to 1000\,km/s along the line of sight, it could introduce a bias in radial distance up to 3.5\,Mpc.}. The distance between \hls\, and HLS-2-1 corresponds to 5.0 comoving Mpc, which is comparable to the scale of the z$\sim$5 over-densities found in COSMOS and GOODS-N associated with SMG/DSFGs \citep{Mitsuhashi+21,Herard-Demanche+23}. When assuming the core of the possible structure has the same redshift as HLS-2-1 and \hls, the deviation of the redshift of HLS-2-2 would correspond to a line-of-sight comoving distance of 58\,Mpc. This is an order of magnitude larger than the scale of the SMG over-density in COSMOS, while still being comparable to the proto-clusters traced by Ly-$\alpha$ emitters at z=5-6 \citep{Jiang+18,Calvi+21}. Although the stochasticity of star formation makes SMG a unreliable tracer of the most massive halos at intermediate redshift, the high SFR of the 3 sources at such high redshift could only be produced by the most massive galaxies tracing the densest environments in the early Universe \citep{Miller+15}. A more complete redshift survey on the other NIKA2 sources in the HLS field, as well as deep optical-IR observation in the same region, could possibly reveal more members of this possible galaxy over-density to confirm its nature and understand its fate of cosmic evolution.

\section{Summary and conclusions}\label{sc:conclusions}

We present the study on 4 DSFGs selected from the early science verification observations of NIKA2, the KIDs camera installed on the IRAM 30m telescope. 

We develop a new framework to determine the redshift of sources with the joint analysis of multi-wavelength photometry and millimeter spectral scans. Accounting for the additional constraints on IR luminosity from the SED modeling, we predict the flux of the strongest emission lines from CO, [CI] and [CII], generate the model spectra at given redshifts accordingly, estimate the goodness of match between the broad-band SEDs, models and the observed millimeter spectra altogether and quantitatively find the most probable redshift solutions based on all this information. 

Based on the prior selection on red far-IR to millimeter colors, we identify a sample of 4 millimeter NIKA2 sources of mJy fluxes in the HLS field with possible high redshifts, at $z=3-7$. We conducted deep NOEMA observations on these sources, and resolve them into 5 individual sources. With the NOEMA spectral scans and the newly developed joint-analysis method, we obtain their redshift and confirm they all have $z>3$. Our analysis reveals that most of their properties, such as star formation rate, dust temperature and gas depletion time are normal compared to typical high-z DSFGs with very active star formation. However, we also find that one of our source (HLS-3) shows significantly low dust temperature and long gas depletion time, resembling the properties of secularly-evolved main sequence star-forming galaxies. Furthermore, we find two sources at z=5.2 that are separated by only 5 comoving Mpc, possibly linked to a third source lying at a distance comparable to the proto-cluster size as traced by Ly-$\alpha$ emitters at $z=5-6$. This could be the hint of an interesting high-$z$ structure in this field. \\

We demonstrate that our method to constrain the redshift, applied to millimeter selected DSFGs with only far-IR to mm photometry and blind spectral scans, could determine the true redshift accurately. Such accuracy of redshift determination with multiple low SNR emission lines shows promising potential in blind redshift searching on large sample of high-z millimeter-faint DSFGs, even in the absence of accurate optical-IR photometric redshifts. The method is especially expected to improve the design and efficiency of blind redshift search on candidate high-$z$ DSFGs detected by the NIKA2 Cosmological Legacy Survey (N2CLS). Indeed, most of N2CLS sources are fainter (sub-mJy) than the 4 sources discussed here. The new tool we developed will allow us to mitigate the increase of NOEMA or ALMA time that will be needed for these faint DSFGs. \\

The joint analysis methods also provide possible implications to the strategy to obtain accurate redshift and cosmic evolution of high-z DSFGs. The next generation single-dish telescopes/instruments, such as the CCAT-prime \citep{CCAT-P+21} and LMT TolTEC \citep{Wilson+20}, are planned to devote a substantial fraction of observing times in wide area deep blind surveys. With thousands of DSFGs expected to be detected, these surveys aim to reveal the role of DSFGs in the formation and evolution of massive galaxies through their cosmic evolution and environment/clustering. However, comparing to the existing deep millimeter surveys, the majority of these planned surveys are not expected to be completely covered by deep surveys in near IR at $>$2$\mu$ \citep{Wang+19,Williams+19,Fudamoto+21,Xiao+23}. The lack of the wide and deep near IR surveys like COSMOS-Web \citep{Casey+22} could make it difficult to identify the counterpart of high-$z$ DSFGs, which further prevent the application of optical-IR SED modeling for efficient and accurate redshift measurements. Our practice on the HLS sources under similar conditions, however, demonstrate that the joint constraints of photometric redshift, IR luminosity and millimeter spectra from far-IR SED and blind spectral scans could also provide a promising accuracy and robustness in efficient redshift searching of high-z DSFGs. Further improvement following this strategy, including the application of this method to the redshift identification of a larger sample of DSFGs discovered by the NIKA2 Cosmological Legacy Survey \citep{Bing+23}, is expected to benefit the key scientific objectives of these future wide area (sub)millimeter surveys.

\begin{acknowledgements}
LB warmly acknowledges financial support from IRAM for his first year of PhD thesis and the support from the China Scholarship Council grant (CSC No. 201906190213).

We acknowledge financial support from the ”Programme National de Cosmologie and Galaxies” (PNCG) funded by CNRS/INSU-IN2P3-INP, CEA and CNES, France, from the European Research Council (ERC) under the European Union's Horizon 2020 research and innovation programme (project CONCERTO, grant agreement No 788212) and from the Excellence Initiative of Aix-Marseille University-A*Midex, a French "Investissements d'Avenir" programme. 

This work is based on observations carried out under project numbers W16EE, E16AI, W17EL, W17FA, W18FA, and S20CL with the IRAM NOEMA Interferometer and project numbers 230-14 and 192-16 with the IRAM 30m telescope. IRAM is supported by INSU/CNRS (France), MPG (Germany) and IGN (Spain).

We would like to thank the IRAM staff for their support during the NIKA and NIKA2 campaigns. The NIKA2 dilution cryostat has been designed and built at the Institut N\'eel. In particular, we acknowledge the crucial contribution of the Cryogenics Group, and in particular Gregory Garde, Henri Rodenas, Jean Paul Leggeri, Philippe Camus. This work has been partially funded by the Foundation Nanoscience Grenoble and the LabEx FOCUS ANR-11-LABX-0013. This work is supported by the French National Research Agency under the contracts "MKIDS", "NIKA" and ANR-15-CE31-0017 and in the framework of the "Investissements d’avenir” program (ANR-15-IDEX-02). This work has benefited from the support of the European Research Council Advanced Grant ORISTARS under the European Union's Seventh Framework Programme (Grant Agreement no. 291294). F.R. acknowledges financial supports provided by NASA through SAO Award Number SV2-82023 issued by the Chandra X-Ray Observatory Center, which is operated by the Smithsonian Astrophysical Observatory for and on behalf of NASA under contract NAS8-03060. 

In addition to the NIKA2 collaboration pipeline, the NIKA2 data were also processed using the Pointing and Imaging In Continuum (PIIC) software, developed by Robert Zylka at the Institut de Radioastronomie Millimetrique (IRAM) and distributed by IRAM via the GILDAS pages. PIIC is the extension of the MOPSIC data reduction software to the case of NIKA2 data.

We warmly thank Véronique Buat for insightful discussions on the results of this paper.

\end{acknowledgements}

%
%

\bibliographystyle{aa} 
\bibliography{NIKA2_SV/NIKA2_SV} 

\begin{appendix} 
%

\section{Robustness of the joint-likelihood method using different line widths}\label{sc:test_linewidth}

One of the key assumption is the width of the emission lines, which we fixed to 500km/s. Previous studies reveal a correlation between total IR/line luminosity and far-IR to millimeter line width, possibly originating from the regulation of gaseous disk rotation by gravity or (and) feedback from star formation or AGN \citep{Bothwell+13,Goto+15}. The assumed line width generally matches the average of DSFGs with ULIRG-HyLIRG luminosities in infrared, which is similar to the derived IR luminosity of our sample. However, observations also show significant scatters among IR luminous DSFGs. Our sample could be a typical example of the variety of line width of luminous DSFGs, which have line FWHMs ranges from $\sim$250km/s (HLS-2-1) to $\sim$750km/s (HLS-3). Besides, the assumption of Gaussian line profile generally holds for most of our source, but HLS-3, as described in Sect.~\ref{sc:properties_gas_hls}, has a significant double-peak feature in the detected emission line in band2.

The impact on the joint analysis result from the mismatch between real and assumed line width and profile are an uncertain prior. Thus, we make the following tests to check if and how the results of this joint analysis could change with the assumption on different line widths. In addition to the default setting of 500\,km/s FWHM Gaussian line profile, we further perform the joint analysis with line profiles FWHMs of  300km/s and 800\,km/s, using the redshift and infrared luminosity derived from the fit with \citet{Bethermin+15}. We identify the redshift solutions of the 4 HLS sources with at least 1 line detected with these 2 different assumptions, using the same method and criteria described as in Sect.~\ref{sc:joint_pdf_z}. The results are listed in Table~\ref{z_HLS_diffvel}. 

\begin{figure*}[tb]
   \centering
      \includegraphics[clip=true,width=1.0\textwidth]{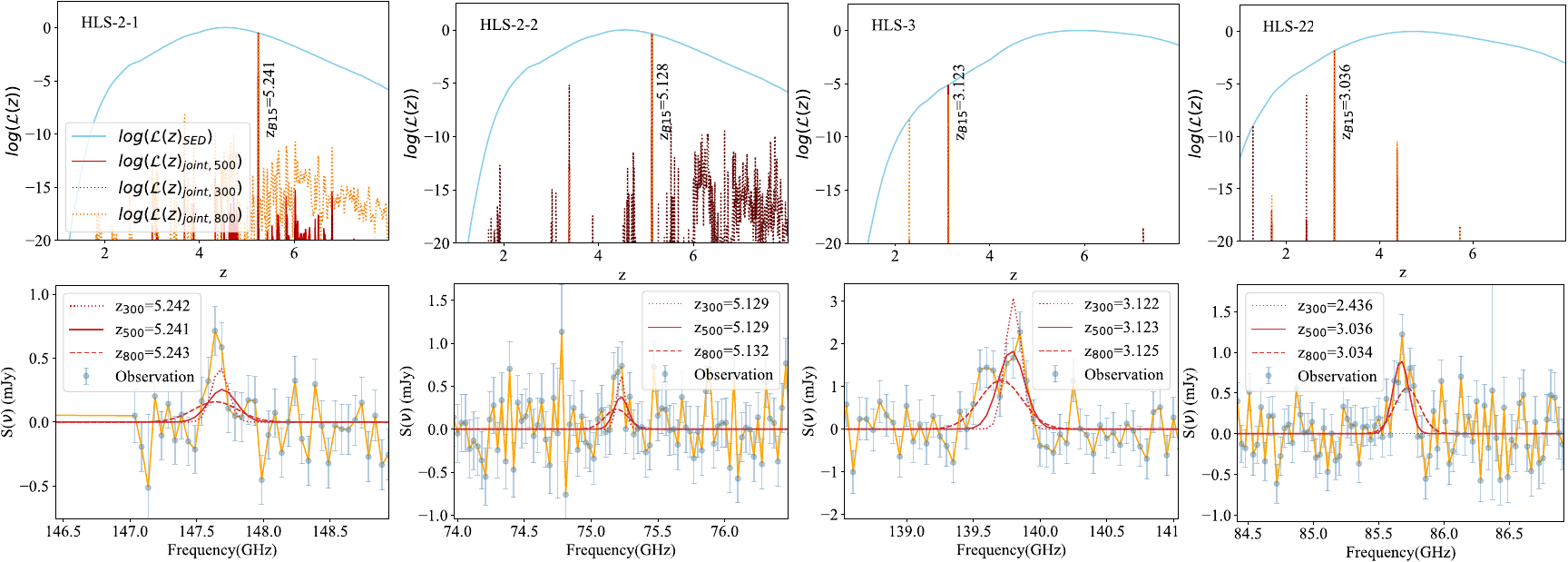}
      \caption{Upper row: Joint log-likelihood of 4 HLS sources with models spectra with line widths of 300km/s, 500km/s and 800km/s. Lower row: Comparison between the models of these 3 line widths at the corresponding redshift solution and the observed source spectra. }
      \label{plot_diffvel}
   \end{figure*}

From the results in Table~\ref{z_HLS_diffvel}, we conclude that the redshift solutions from the joint analysis method are generally not sensitive to the assumptions on emission line widths. For all sources but HLS-22, we find little variation in the redshift solutions using different line width assumptions. The differences of $\Delta z\sim 0.003$, as shown in Fig.\,\ref{plot_diffvel}, are mostly originating from the changes of the peak intensity of emission lines, which could lead to slight variations of $\chi^2(z)$. However, such little difference is still within the width of the emission lines, and will neither cause false identification of emission line, nor affect the analysis on line fluxes and kinematics in Sect.\,\ref{sc:properties_gas_hls} with the corresponding central frequency as an initial guess. 

\begin{table}[h]
\caption{Redshift of HLS sources from the joint-analysis with models of different line width}           
\label{z_HLS_diffvel}   
\centering
\begin{tabular}{c c c c}       
\hline\hline                 
Name & z$_{500 km/s}$ & z$_{300 km/s}$ & z$_{800 km/s}$  \\    
\hline                                   
   HLS-2-1 & 5.241 & 5.242 & 5.243 \\ 
   HLS-2-2 & 5.129 & 5.129 & 5.132 \\
   HLS-3   & 3.123 & 3.122 & 3.125 \\   
   HLS-22  & 3.036 & 2.436 & 3.034 \\ \hline          
   \end{tabular}
\end{table}

\begin{figure}[tb]
   \centering
      \includegraphics[clip=true,width=0.5\textwidth]{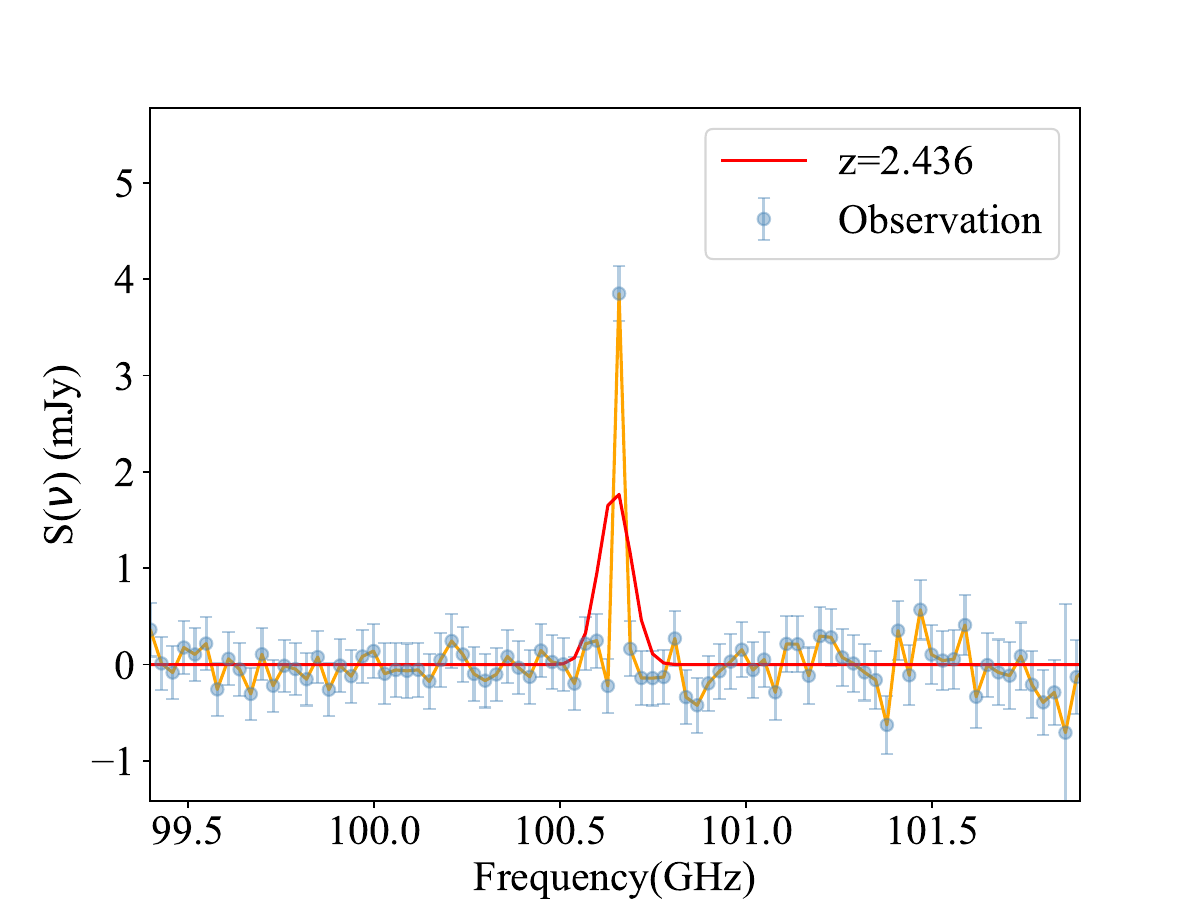}
      \caption{Comparison between the observed spectra of HLS-22 and the model spectra with FHWM of 300\,km/s at the best redshift solution z=2.436. The feature identified as an emission line might be a missing glitch within the data.}
      \label{plot_spike_HLS22}
   \end{figure}

The only case of significant inconsistency in redshift from the test is HLS-22, where the procedure using 300km/s line width strongly favors a redshift solution at $z=2.436$. Given the frequency of the 2 emission lines detected with high SNR, we are confident about the redshift solution at $z=3.036$ from the analysis with the model spectra of 500km/s line width. Therefore, We checked the 300km/s model and the data at z=2.436 and find that the mis-identification is caused by a strong noise spike at 100.64\,GHz, as shown in Fig.~\ref{plot_spike_HLS22}. The false identification suggests an increased sensitivity to narrow spikes in the spectrum with the decrease of model line width. In our calculation of $\chi^2_{spec}(z)$ and correspondingly, joint log-likelihood, their variation with redshift are dominated by the goodness of match between model and data within the range of model line profiles. With a narrower line width in the model, the number of data points that dominate the variation of $\chi^2_{spec}$(z) will be smaller compared to the cases with wider line width. This will make the analysis with narrow line width more sensitive to single spurious data points, like the noise spike in HLS-22 spectra, and lead to the mis-identification in Fig.\,\ref{plot_spike_HLS22}. A less aggressive spectral binning along frequency and a pre-processing with sigma clipping could probably reduce such false identification in practice. 

\section{Robustness of the joint-likelihood method with narrower frequency coverage}\label{sc:test_specoverage}

\begin{figure*}[tb]
   \centering
      \includegraphics[clip=true,width=1.0\textwidth]{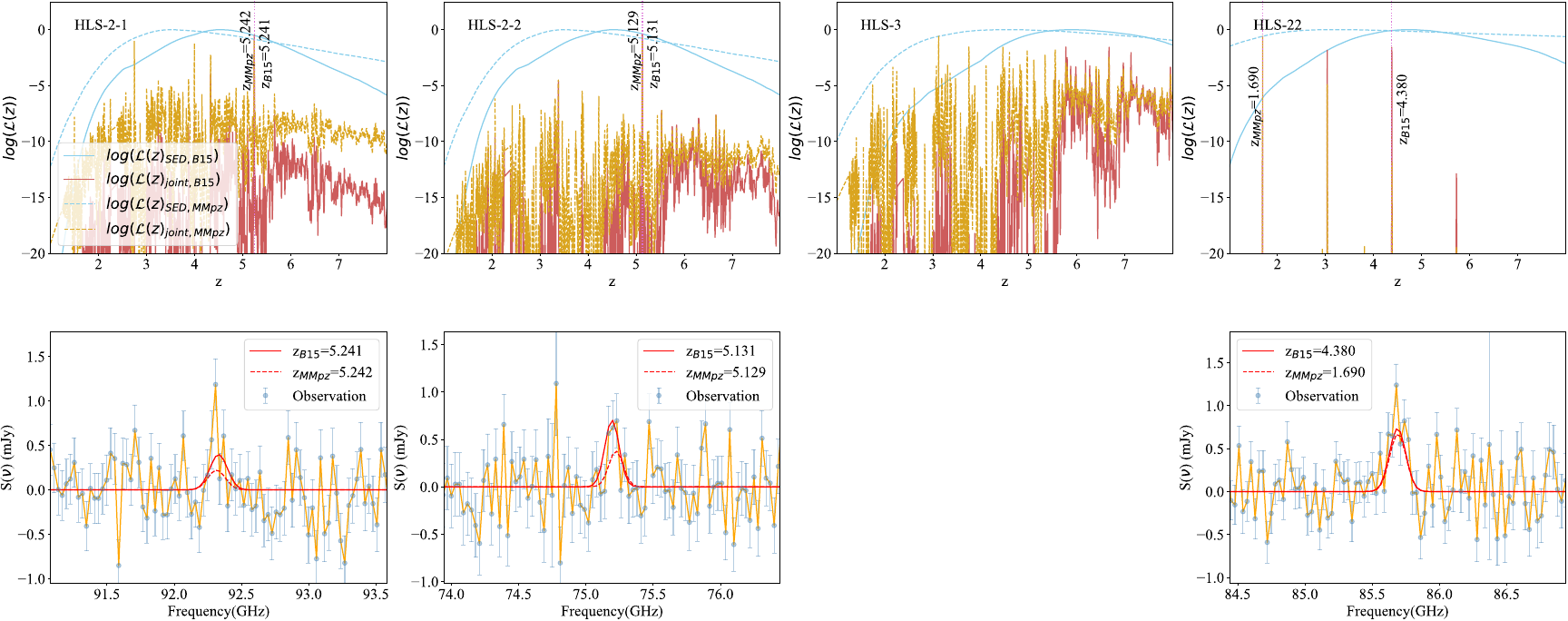}
      \caption{The result of joint-analysis on the 4 HLS sources with spectroscopic redshifts derived in Sect.~\ref{sc:joint_pdf_z}, using only the 31\,GHz NOEMA spectral scans observed in 2018. First row shows the likelihood from SED fittings and joint log-likelihood of photometric and spectroscopic data, using the SED fitting outputs with \citet{Bethermin+15} templates and MMPZ. The comparison between observed spectra and the model spectra predicted by the IR luminosities from the 2 SED fitting results are shown in the second and third row.}
      \label{joint_hls_2018}
   \end{figure*}

Millimeter spectral scans made by interferometers are widely used to blindly search for the emission lines from candidate high redshift DSFGs, determine their spectroscopic redshift and study the conditions of their cold ISM \citep{Strandet+16,Fudamoto+17,Jin+19,Neri+20,Reuter+20}. These spectral scan observations are designed to cover a continuous frequency range with several spectral setups. For ALMA and NOEMA, the default setup of the blind spectral scans at their current lowest frequency band covers $\sim$ 31\,GHz. The earliest observations in 2018 on HLS sources blindly and continuously cover the spectra of HLS sources between 71\,GHz and 102\,GHz in NOEMA band1 with 2 setups, which follows this basic strategy of blind redshift search. To test the joint analysis method under more realistic conditions in large DSFG redshift survey projects, we apply the method to analyse these band1 spectra and compare their resulting redshifts with the ones in Sect.~\ref{sc:joint_pdf_z}. The results from this analysis are presented in Table~\ref{z_HLS_spec2018} and Fig.~\ref{joint_hls_2018}.

\begin{table}
\caption{Redshift of HLS sources from the joint-analysis with only the 31\,GHz continuous spectra observed in 2018.}           
\label{z_HLS_spec2018}   
\centering
\begin{tabular}{c c c c}       
\hline\hline                 
Name & z$_{true}$ & z$_{B15,2018}$ & z$_{MMPZ,2018}$  \\    
\hline                                   
   HLS-2-1 & 5.241 & 5.241 & 5.242 \\ 
   HLS-2-2 & 5.129 & 5.131 & 5.129 \\
   HLS-3   & 3.123 & N/A & N/A \\   
   HLS-22  & 3.036 & 4.380 & 1.690 \\ \hline                                  
   \end{tabular}
\end{table}

From the comparison between the redshift analysis using early and full datasets, we find that the best redshift solutions of HLS-2-1 and HLS-2-2 remain stable, while the results of HLS-22 and HLS-3 are affected by the narrowed spectral coverage. Such a difference in robustness under different spectral coverage could be explained as follows. The joint analysis method works equivalently to the automatic alignment and stacking of 2 or more lines in the spectra. If it is at the correct redshift and with multiple lines covered, this method could numerically boost the stacked SNR of emission lines, even if none of the single lines are detected with high significance. At a fixed coverage in frequency, we could expect that sources with higher redshift could have more CO lines to be covered. Taking our sample as an example, although the lines of HLS-2-1 and HLS-2-2 at z$\sim$ 5.2 are only tentatively detected, their relatively high redshifts ensure that at least two CO/[CI] lines are covered by the spectral scan. On the contrary, the narrowed spectral coverage leaves only one CO line in the spectral coverage of early observations of HLS-3 and HLS-22, leading to ambiguous redshift solutions no matter if the line is detected at high significance. With the comparison of the redshift robustness of these two groups of sources under different spectral coverage, we also emphasize that wide spectral coverage covering at least two strong molecular/atomic lines could be even more crucial in the redshift identification of DSFGs compared to reaching high sensitivity.

\section{Cross-validation and tension between different SED modeling}\label{sc:test_consistency}

The application of the joint-analysis framework on HLS sources largely relies on the current knowledge on the far-IR SED of high-z galaxies. However, although Herschel provides estimates on the mean far-IR SEDs and the redshift evolution of the main population of star-forming galaxies, these results are also limited by significant source confusion, especially in SPIRE data at longer wavelengths. Moreover, current studies reveal some DSFGs with apparently low-dust temperature \citep{Jin+19}, as well as a significant warm-dust contribution in some starburst galaxies \citep{Eisenhardt+12,Wu+12,Fan+16}. These results suggest that a large variation in far-IR SEDs could exist in high-z DSFG populations.

Our choices of template could not be free from these issues, and this is the reason why we adjust our joint-analysis framework to the results of 2 different far-IR SED templates and modeling framework, and make the cross-validation between the results of the two. The analysis with \citet{Bethermin+15} templates and MMPZ mostly shows consistent redshift solutions. This suggests the relative stability of the joint-analysis method with input information from different SED fitting results.

However, some discrepancies on HLS-22 when using typical blind spectral scan conditions in Appendix\,\ref{sc:test_specoverage} are also found, which leads us to have an additional check on its origin. From the comparison on derived IR luminosity in Fig.~\ref{hls_redshift}, and the comparison between model and data in Fig.~\ref{joint_redshift_HLS} and Fig.~\ref{joint_hls_2018}, we notice that the estimated infrared luminosity and line fluxes from MMPZ are systematically lower than those from \citet{Bethermin+15}. At the correct redshift, the predicted line fluxes of \citet{Bethermin+15} match better with the observed line fluxes compare to MMPZ. On the contrary, MMPZ generally returns more accurate photometric redshifts, especially on HLS-3, where the photometric redshift from \citet{Bethermin+15} significantly deviates from the spectroscopic redshift. However, as indicated by the low dust temperatures of the HLS sample, it is possible that the properties of far-IR emission of these galaxies are not representative among high redshift star-forming galaxies. Thus, we decide not to make any preference on the choice of dust template and far-IR SED fitting in our framework, and we recommend a cross-validation between the redshift solutions from various method in application.

Besides, the faint emission of HLS sources in SPIRE bands introduce large uncertainties on the constraints on source SEDs around the peak of far-IR emission. This, as a result, could contribute to the difference in IR luminosities derived from methods with different prior constraints \citep{Casey+20}. These issues also further suggest the importance of matching observation at ALMA band 8-10 frequencies in properly reconstructing the far-IR SED, as well as estimating the IR luminosity and star formation rate of high-z DSFGs selected by millimeter surveys.

\section{Impact of the Scatter of L$_{FIR}$-L$_{CO}$ Scaling Relation}\label{sc:test_lfir_lco}

The redshift from the joint analysis is derived based on the goodness of match between the emission line model and observed spectra. As mentioned in Sect.~\ref{sc:joint_pdf_z}, in this approach we predict the expected fluxes of spectral lines based on the L$_{FIR}$ from the SED template fitting using the best-fit scaling relation between L$_{FIR}$-L$_{line}$ from literature \citep{Greve+14, Liu+15, Valentino+18}. However, these scaling relations are subject to substantial scatter up to a factor of a few in observations. To test the possible impact of these scatters on our analysis and the robustness of the joint analysis method against them, we first checked the output best redshift solution after adding a systematic offset to all L$_{FIR}$-L$_{line}$ scaling relations when generating the predicted spectra models. In this test, we shift the predicted CO/[CI] line fluxes by four different systematical offsets corresponding to $\pm$0.5 and $\pm$1.0 times of the 1$\sigma$ scatter of the scaling relations. The exact values of the 1$\sigma$ scatters for the considered lines are given in Table.~\ref{scaling_relation}.

The best redshift solutions for HLS sources (except for HLS-4) after applying these four different offsets in L$_{FIR}$-L$_{line}$ conversion are listed in Table~\ref{z_HLS_offset_lirlco}. We find that all offsets in line flux, but the +1.0$\sigma$, result in best redshift solutions similar to the analysis using the median (zero offset) scaling relations. This suggests good robustness of our joint analysis method against the existing scatter of L$_{FIR}$-L$_{line}$ scaling relations in observations. As for the test with +1.0$\sigma$ offset, we further checked the reason leading to the discrepancy in best redshift solution. For HLS-22 and HLS-2-2, the application of the +1.0$\sigma$ offset leads to mismatches between the model and the data due to glitches or noise spikes in the spectra (see Fig.~\ref{plot_spike_HLS22} as an example, where the noise spike is matched with CO(5-4) at the best redshift solution of HLS-22 here). For HLS-3, we find this unlikely low redshift solution after applying the +1.0$\sigma$ offset as the code assigns the only strong emission line in the spectral scan at 139.746\,GHz to be CO(2-1). This suggests the stronger demand of having spectral scan wide enough to cover more than one strong spectral line in the redshift confirmation of DSFGs with moderate redshift (i.e z$\sim$2-3).

\begin{table}[h]
\caption{Redshift of HLS sources from the joint-analysis when using different amount of offsets for all L$_{FIR}$-L$_{line}$ scaling relations.}           
\label{z_HLS_offset_lirlco}   
\centering
\begin{tabular}{c c c c c c}       
\hline\hline                 
Name & z$_\mathrm{best,med}$ & z$_\mathrm{best,-1.0\sigma}$ & z$_\mathrm{best,-0.5\sigma}$ & z$_\mathrm{+0.5\sigma}$ & z$_\mathrm{+1.0\sigma}$ \\    
\hline                                   
   HLS-2-1 & 5.241 & 5.241 & 5.241 & 5.241 & 5.241 \\ 
   HLS-2-2 & 5.129 & 5.128 & 5.128 & 5.128 & 6.305 \\ 
   HLS-3   & 3.123 & 3.123 & 3.123 & 3.122 & 0.649 \\   
   HLS-22  & 3.036 & 3.036 & 3.036 & 3.036 & 5.724 \\  \hline          
   \end{tabular}
\end{table}

\begin{figure*}[tb]
   \centering
      \includegraphics[clip=true,width=1.0\textwidth]{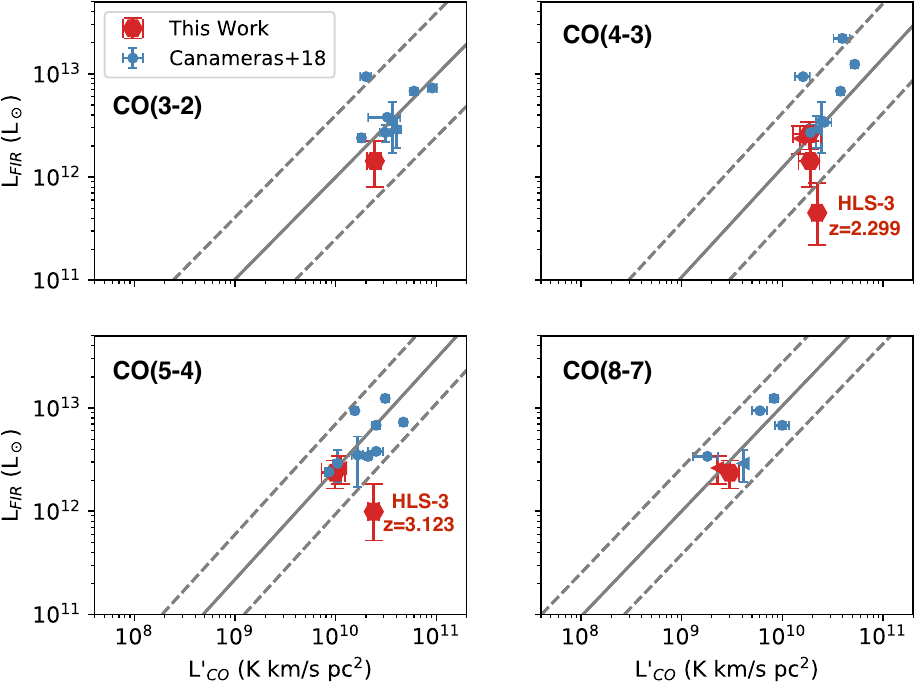}
      \caption{The comparison between L$_\mathrm{FIR}$-L$_\mathrm{CO}$ correlations of different transitions \citep{Greve+14,Liu+15} and our sample based on measurements from our observations at the best redshift solutions. The sources with upper limits on line luminosities are presented as leftward triangles.}
      \label{lir_lco_check}
   \end{figure*}

To test the self-consistency of the joint analysis method, we put the measured CO line fluxes and far-IR luminosities of HLS sources at their best redshift solutions (the z$_\mathrm{best,med}$ in Table~\ref{z_HLS_offset_lirlco}) in the corresponding L$_\mathrm{FIR}$-L$_\mathrm{CO}$ diagrams to check if they follow the scaling relations used for line flux predictions. The results are shown in Fig.~\ref{lir_lco_check}. The plotted CO and far-IR luminosities are derived using the Gaussian-fitted line fluxes and modified black-body fitting (see Table.~\ref{linemeasure} and Table.~\ref{MBB_dust_properties}). Apart from the average L$_\mathrm{FIR}$-L$_\mathrm{CO}$ correlations, we also show the samples from \citet{Canameras+18} with multiple line transitions from the same sources as a comparison to our sources.

From the Fig.~\ref{lir_lco_check}, we find that most of our sources fall within +-2$\sigma$ of the scaling relation used in our analysis, which is also consistent with the regions occupied by bright sub-millimetre galaxies in \citet{Canameras+18}. The only exception is HLS-3, which is also highlighted in Fig.~\ref{lir_lco_check}. On either L$_{FIR}$-L$^{'}_{CO(5-4)}$ or L$_{FIR}$-L$^{'}_{CO(4-3)}$ diagram, HLS-3 falls well below the scaling relation even if we consider the scatter of these scaling relations. However, we also note that it has one of the poorest SPIRE photometry among all of the four HLS sources. As the SPIRE bands close to the peak wavelength of SED predominantly constrain the IR luminosity, it is likely that the IR luminosity of HLS-3 is much less constrained than the rest HLS sources, especially compared to HLS-2 and HLS-4.

\end{appendix}

\end{document}